\documentstyle[agupp,psfig]{article} 
\lefthead{HELMSTETTER,  SORNETTE, GRASSO and OUILLON}
\righthead{Mainshocks are Aftershocks of Conditional Foreshocks} 
\authoraddr{Agn\`es Helmstetter, Laboratoire de G{\'e}ophysique
Interne et Tectonophysique,
Observatoire de Grenoble, Universit{\'e} Joseph Fourier, BP 53X,
38041 Grenoble Cedex, France. (e-mail: ahelmste@obs.ujf-grenoble.fr)}

\authoraddr{Didier Sornette, Department of Earth and Space
Sciences and Institute of Geophysics and Planetary Physics, University
of California, Los Angeles, California and Laboratoire de Physique de
la Mati\`{e}re
Condens\'{e}e, CNRS UMR 6622 Universit\'{e} de Nice-Sophia Antipolis,
Parc Valrose, 06108 Nice, France  (e-mail: sornette@ess.ucla.edu)}

\authoraddr{Jean-Robert Grasso, Laboratoire de G{\'e}ophysique
Interne et Tectonophysique,
Observatoire de Grenoble, Universit{\'e} Joseph Fourier, BP 53X,
38041  Grenoble Cedex, France. (e-mail:Jean-Robert.Grasso@obs.ujf-grenoble.fr)}

\input{update.tex}
\begin{document}
\title{Mainshocks are Aftershocks of Conditional Foreshocks:\\
How do Foreshock Statistical Properties Emerge from Aftershock Laws}
\author{Agn\`es Helmstetter}
\affil{Laboratoire de G{\'e}ophysique Interne et Tectonophysique,
         Observatoire de Grenoble, Universit\'e Joseph Fourier, France}
\author{Didier Sornette}
\affil{ Laboratoire de Physique de la Mati\`{e}re Condens\'{e}e, CNRS
UMR 6622
Universit\'{e} de Nice-Sophia Antipolis, Parc Valrose, 06108 Nice,
France and Department of Earth and Space Sciences and Institute of
Geophysics and Planetary Physics, University of California, Los
Angeles, California 90095-1567}
\author{Jean-Robert Grasso}
\affil{Laboratoire de G{\'e}ophysique Interne et Tectonophysique,
         Observatoire de Grenoble, Universit\'e Joseph Fourier, France}


\newcommand{\be}{\begin{equation}}
\newcommand{\ee}{\end{equation}}
\newcommand{\ba}{\begin{eqnarray}}
\newcommand{\ea}{\end{eqnarray}}
\newenvironment{technical}{\begin{quotation}\small}{\end{quotation}}

\begin{abstract}
The inverse Omori law for foreshocks discovered in the 1970s
states that the rate of earthquakes prior to a mainshock increases
on average as a power law $\propto 1/(t_c-t)^{p'}$ of the time to the
mainshock occurring at $t_c$.
Here, we show that this law results from the direct Omori law for
aftershocks describing the power law decay $\sim 1/(t-t_c)^{p}$ of seismicity
after an earthquake, provided that any earthquake can trigger its
suit of aftershocks. In this picture, the seismic activity at any time is
the sum of the spontaneous tectonic loading and of the
activity triggered by all preceding events weighted by their
corresponding Omori law. The inverse Omori law then emerges as the
expected (in a statistical sense) trajectory of seismicity, conditioned
on the fact that it leads to the burst of seismic activity
accompanying the mainshock.
In particular, we predict and verify by numerical simulations
on the Epidemic-Type-Aftershock Sequence (ETAS) model that $p'$ is always
smaller than or equal to $p$ and a function of $p$, of the $b$-value of the
Gutenberg-Richter law (GR) and of a parameter quantifying the number
of direct aftershocks as a function of the magnitude of the mainshock.
The often documented apparent decrease of the $b$-value
of the GR law at the approach to the main
shock results straightforwardly from the conditioning of the path of
seismic activity culminating at the mainshock. However, we predict that the
GR law is not modified simply by a change of $b$-value but that a more accurate
statement is that the GR law gets an additive (or deviatoric) power law
contribution with exponent smaller than $b$ and with an amplitude
growing as a power law of the time to the mainshock.
In the space domain, we predict that the phenomenon of aftershock diffusion
must have its mirror process
reflected into an inward migration of foreshocks towards the mainshock.
In this model, foreshock sequences are special aftershock
sequences which are modified by the condition to end up in a burst of
seismicity associated with the mainshock. Foreshocks are
not just statistical creatures, they are genuine forerunners of large shocks
as shown by the large prediction gains obtained using several of
their qualifiers.
\end{abstract}

\begin{article}

\section{Introduction}

Large shallow earthquakes are followed by an increase in
seismic activity,
defined as an aftershock sequence.
It is also well-known that large earthquakes are sometimes
   preceded by an
unusually large activity rate, defined as a foreshock sequence.
Omori law describing the power law decay $\sim 1/(t-t_c)^{p}$
of aftershock rate with time from a mainshock that occurred at $t_c$
has been proposed more than one century ago [{\it Omori}, 1894], and has
since been verified by many studies [{\it Kagan and Knopoff}, 1978;
{\it Davis and Frohlich}, 1991;
{\it Kisslinger and Jones}, 1991; {\it Utsu et al.}, 1995]. See however
[{\it Kisslinger}, 1993; {\it Gross and Kisslinger}, 1994] for alternative
decay laws such as the stretched exponential and its possible explanation
[{\it Helmstetter and Sornette}, 2002a].

Whereas the Omori law describing the aftershock decay rate is one of
the few well-established empirical laws in seismology, the increase of
foreshock rate before an
earthquake does not follow such a well-defined empirical law.
There are huge fluctuations of the foreshock seismicity rate, if any, from one
sequence of earthquakes to another one preceding a mainshock.
Moreover, the number of foreshocks per mainshock is usually quite smaller than
the number of aftershocks. It is thus essentially impossible to establish
a deterministic empirical law that describes the intermittent
increase of seismic
activity prior to a mainshock when looking at a single foreshock sequence which
contains at best a few events.
Although well-developed individual foreshock sequences are rare and
mostly irregular,
a well-defined acceleration of foreshock rate prior to a mainshock emerges
when using a superposed epoch analysis, in other words, by synchronizing
several foreshock sequences to a common origin of time defined as the time
of their mainshocks and by stacking these synchronized foreshock sequences.
In this case, the acceleration of the seismicity  preceding the mainshock
clearly follows an inverse Omori law of the form $N(t) \sim 1/(t_c-t)^{p'}$,
where $t_c$ is the time of the mainshock.
This law  has been first proposed by {\it Papazachos} [1973], and has been
established more firmly by [{\it Kagan and Knopoff}, 1978; {\it Jones
and Molnar}, 1979]. The inverse Omori law is usually observed for time scales 
smaller than the direct Omori law, of the order of weeks to months before 
the mainshock.

A clear identification of foreshocks, aftershocks and mainshocks is hindered
by the difficulties in associating an unambiguous and unique
space-time-magnitude domain to any earthquake sequence.
Identifying aftershocks and foreshocks requires the definition of a space-time
window. All events in the same space-time domain define a sequence.
The largest earthquake in the sequence is called the mainshock.
The following events  are identified as aftershocks, and the preceding events
are called foreshocks.

Large aftershocks show the existence of secondary
aftershock activities, that is, the fact that aftershocks may have
their own aftershocks, such as the $M=6.5$ Big Bear event, which is
considered as an
aftershock of the $M=7.2$ Landers Californian earthquake, and which clearly
triggered its own aftershocks.
Of course, the aftershocks of aftershocks can be clearly identified
without further insight and analysis as obvious bursts of transient
seismic activity
above the background seismicity level, only for the largest aftershocks.
But because aftershocks exist on all scales, from the laboratory scale, 
e.g. [{\it Mogi}, 1967; {\it Scholz}, 1968], to the worldwide seismicity,
we may expect that all earthquakes, whatever their magnitude,
trigger their own aftershocks, but with a rate increasing with the
mainshock magnitude, so that only aftershocks of the largest earthquakes
are identifiable unambiguously.

The properties of aftershock and foreshock sequences depend on the
choice of these space-time windows, and on the specific definition of
foreshocks [e.g. {\it Ogata et al.}, 1996], which can
sometimes be rather arbitrary. In the sequel, we shall consider two
definitions of foreshocks for a given space and time window:
\begin{enumerate}
\item  we shall call ``foreshock'' of type I any event of magnitude smaller
than or equal to the magnitude of the following event, then identified
as a ``main shock''. This definition implies the choice of a space-time
window $R \times T T$ used to define both foreshocks and mainshocks. 
Mainshocks are
large earthquakes that were not preceded by a larger event in this
space-time window. The same window is used to select foreshocks
before mainshocks;

\item we shall also consider ``foreshock''  of type II, as any earthquake
preceding a large earthquake, defined as the mainshock,  
 independently of the relative magnitude
of the foreshock compared to that of the mainshock. This second
definition will thus incorporate seismic sequences in which a foreshock
could have a magnitude larger than the mainshock, a situation which can
alternatively be interpreted as a mainshock followed by a large aftershock.
\end{enumerate}
The advantage of this second definition is that foreshocks of type II are
automatically defined as soon as one has identified the mainshocks,
for instance, by calling mainshocks all events of magnitudes larger than
some threshold of interest.
Foreshocks of type II are thus all events preceding these large
magnitude mainshocks. In contrast, foreshocks of type I need to obey a
constraint on their magnitude, which may be artificial, as suggested
from the previous discussion.
All studies published in the literature deal with foreshocks of type I.
Using a very simple model of seismicity, the so-called ETAS (epidemic-type
aftershock) model, we shall show that the definition of foreshocks of
type II is also quite meaningful and provides new insights for classifying
earthquake phenomenology and understanding earthquake clustering in
time and space.

The exponent $p'$ of  the inverse Omori law is usually found to be
smaller than or close to $1$ [{\it Papazachos et al.}, 1967;
{\it Papazachos et al.}, 1975b;
{\it Kagan and Knopoff}, 1978; {\it Jones and Molnar}, 1979; {\it 
Davis and Frohlich},
1991; {\it Shaw}, 1993; {\it Ogata et al.}, 1995; {\it Maeda}, 1999;
{\it Reasenberg}, 1999],
and is always found smaller than or equal to the direct Omori exponent $p$
when the 2 exponents $p$ and $p'$ are measured simultaneously on the same
mainshocks [{\it Kagan and Knopoff}, 1978; {\it Davis and
Frohlich}, 1991;
{\it Shaw}, 1993; {\it Maeda}, 1999; {\it Reasenberg}, 1999].
{\it Shaw} [1993] suggested in a peculiar case the relationship $p'=2p-1$,
based on a clever  but slightly incorrect reasoning (see below).
We shall recover below this relationship only in a certain regime
of the ETAS model from an exact treatment of the
foreshocks of type II within the framework of the ETAS model.

Other studies tried to fit a power law increase of seismicity to individual
foreshock sequences. Rather than the number of foreshocks, these studies
usually fit the cumulative Benioff strain release $\epsilon$ by a power-law
$\epsilon(t) = \epsilon_c - B (t_c-t)^{z}$
with an exponent $z$ that is often found close to $0.3$
(see [{\it Jaum\'e and Sykes}, 1999; {\it Sammis and Sornette}, 2002] 
for reviews).
Assuming a constant Gutenberg Richter $b$-value through time,
so that the acceleration of the cumulative Benioff strain before
the mainshock is due only to the increase in the seismicity rate,
this would argue for a $p'$-value close to $0.7$.
These studies were often  motivated by the critical point theory
[{\it Sornette and Sammis}, 1995], which predicts
a power-law increase of seismic activity before major earthquakes (see e.g.
[{\it Sammis and Sornette}, 2002] for a review).
However, the statistical significance of such a power-acceleration of energy
before individual mainshock is still controversial [{\it Z\"oller and
Hainzl}, 2002].

The frequency-size distribution of foreshocks has also been observed either
to be different from that of aftershocks, $b' < b$, e.g. [{\it Suyehiro}, 1966;
{\it Papazachos et al.}, 1967; {\it Ikegami}, 1967; {\it Berg}, 1968], or to
change as the mainshock is approached. This change of magnitude distribution is
often interpreted as a decrease of $b$-value, first reported by
[{\it Kagan and Knopoff}, 1978; {\it Li et al.}, 1978; {\it Wu et al.}, 1978].
Others studies suggest that the modification of the magnitude distribution is
due only to  moderate or large events, whereas the distribution of
small magnitude
events is not modified [{\it Rotwain et al.}, 1997; {\it Jaum\'e and
Sykes}, 1999].
   {\it Knopoff et al.} [1982] state that only in the rare cases of catalogs
of great length, statistically significant  smaller $b$-value
for foreshocks than for aftershocks are found. Nevertheless they
believe the effect is
likely to be real in most catalogs, but at a very low level of difference.

On the theoretical front, there have been several
models developed to account for foreshocks. Because foreshocks
are rare and seem the forerunners of large events,
a natural approach is to search for physical mechanisms that may explain
their specificity. And, if there is a specificity, this might lead
to the use of foreshocks as precursory patterns for earthquake prediction.
Foreshocks  may result from a slow sub-critical weakening by stress corrosion
[{\it Yamashita and Knopoff}, 1989, 1992; {\it Shaw}, 1993]
or from a general damage process [{\it Sornette et al.}, 1992].
The same mechanism can also reproduce aftershock behavior
    [{\it Yamashita and Knopoff}, 1987; {\it Shaw}, 1993].
Foreshocks and aftershocks may result also from the dynamics of
stress distribution on pre-existing hierarchical structures of faults
or tectonic blocks [{\it Huang et al.}, 1998; {\it Gabrielov et al.}, 2000a,b;
{\it Narteau et al.}, 2000], when assuming that the scale over which stress
redistribution occurs is controlled by the level of the hierarchy (cell size
in a hierarchical cellular automaton model). {\it Dodge et al.}
[1996] argue that
foreshocks are a byproduct of an aseismic nucleation process of a mainshock.
Other possible mechanisms for both aftershocks and foreshocks are based on the visco-elastic response of the crust and on delayed transfer of
fluids in and out of fault structures [{\it Hainzl et al.}, 1999;
{\it Pelletier}, 2000].

Therefore, most of these models suggest a link between aftershocks
and foreshocks.
In the present work, we explore this question further by asking the following
question: is it possible to derive most if not all of the observed
phenomenology
of foreshocks from the knowledge of only the most basic and robust
facts of earthquake
phenomenology, namely the Gutenberg-Richter and Omori laws?
To address this question, we use what is maybe the simplest statistical
model of seismicity, the so-called ETAS (epidemic-type aftershock) model,
based only on the Gutenberg-Richter and Omori laws.
This model assumes that each earthquake can trigger aftershocks, with a rate
increasing as a power law $E^a$ with the mainshock energy $E$, and which decays
with
the time from the mainshock according to the ``local'' Omori law
$\sim 1/(t-t_c)^{1+\theta}$, with $\theta \ge 0$. We stress that the exponent
$1+\theta$ is in general different from the observable $p$-value, as
we shall explain below.
In this model, the seismicity rate is the result of the whole cascade
of direct and secondary aftershocks, that is, aftershocks of aftershocks,
aftershocks of aftershocks of aftershocks, and so on.

In two previous studies of this model, we have analyzed the
super-critical regime
[{\it Helmstetter and Sornette}, 2002a]
and the singular regime [{\it Sornette and Helmstetter}, 2002]
of the ETAS model and have shown that these regimes
can produce respectively an exponential or a power law acceleration of
the seismicity rate. These results can reproduce an individual accelerating
foreshock sequence, but they cannot model the stationary seismicity with
alternative increasing and decreasing seismicity rate before and
after a large earthquake.
In this study, we analyze the stationary sub-critical regime of this branching
model and we show that foreshock sequences are special aftershock
sequences which are modified by the condition to end up in a burst of
seismicity associated with the mainshock. Using only the
physics of aftershocks, all the foreshock phenomenology is derived
analytically and verified accurately by our numerical simulations.
This is related to but fundamentally different from the proposal by
{\it Jones et al.} [1999] that foreshocks are mainshocks whose aftershocks
happen to be big.

Our analytical and numerical investigation of the ETAS model 
gives the main following results:
\begin{itemize}
\item In the ETAS model, the rate of foreshocks increases before the 
mainshock according to the inverse Omori law $N(t) \sim 1/(t_c-t)^{p'}$ 
with an exponent $p'$ smaller than the exponent $p$ of the direct Omori law.
The exponent $p'$ depends on the local Omori exponent $1+\theta$, on the
exponent $\beta$ of the energy distribution, and on the exponent $a$ which
describes the increase in the number of aftershocks with the mainshock energy.
In contrast with the direct Omori law, which is clearly observed after
all large earthquakes, the inverse Omori law is a
statistical law,
which is observed only when stacking many foreshock sequences.

\item While the number of aftershocks increases as the power $E^a$ of
the mainshock
energy $E$, the number of foreshocks of type II is independent of $E$.
Thus, the seismicity generated by the ETAS model increases on average 
according to the inverse Omori law before any earthquake, whatever its 
magnitude. For foreshocks of type I, the same results hold for large 
mainshocks while the conditioning on foreshocks of type I to be smaller 
than their mainshock makes their number increase with $E$ for small and 
intermediate values of the mainshock size.

\item Conditioned on the fact that a foreshock sequence leads to a burst
of seismic activity accompanying the mainshock, we find that the foreshock
energy distribution is modified upon the approach of the mainshock,
and develops a bump in its tail.
This result may explain both the often reported decrease in measured
$b$-value before
large earthquakes and the smaller $b$-value obtained for foreshocks
compared with other earthquakes.

\item In the ETAS model, the modification of the Gutenberg-Richter 
distribution for foreshocks is shown analytically to take the shape of an 
additive correction to the standard power law, in which the new term is 
another power law with exponent $\beta-a$.
The amplitude of this additive power law term also exhibits a kind of inverse
Omori law acceleration upon the approach to the mainshock, with
a different exponent.
These predictions are accurately substantiated by our numerical simulations.

\item When looking at the spatial distribution of foreshocks in the ETAS model,
we find that the foreshocks migrate towards the mainshock as the time increase. 
This migration is driven by the same mechanism underlying
the aftershock diffusion [{\it Helmstetter and Sornette}, 2002b].
\end{itemize}
Thus, the ETAS model, which is commonly used to describe aftershock activity,
seems sufficient to explain the main properties of foreshock behavior in 
the real seismicity.
Our presentation is organized as follows. In the next section, we define
the ETAS model, recall how the average rate of seismicity can be obtained
formally from a Master equation and describe how to deal with fluctuations
decorating the average rate. The third section provides the full
derivation of the inverse Omori law, first starting with an intuitive
presentation followed by a more technical description.
Section four contains the
derivation of the modification of the distribution of foreshock
energies. Section 5 describes the migration of foreshock activity.
Section 6 is a discussion of how our analytical
and numerical results allows us to rationalize previous empirical observations.
In particular, we show that foreshocks are
not just statistical creatures but are genuine forerunners of large
shocks that can be used to obtain significant prediction gains.
Section 7 concludes.

\section{Definition of the ETAS model and its master equation for the
renormalized Omori law}

\subsection{Definitions}

The ETAS model was introduced by {\it Kagan and Knopoff} [1981, 1987] and
{\it Ogata} [1988] to describe the temporal and spatial clustering of 
seismicity
and has since been used by many other workers with success to describe
real seismicity.
Its value stems from the remarkable simplicity of its premises and the small
number of assumptions, and from the fact that it has been shown to fit well
to seismicity data [{\it Ogata}, 1988].

Contrary to the usual definition of aftershocks, the ETAS model does not impose
an aftershock to have an energy smaller than the mainshock.
This way, the same underlying physical mechanism is assumed to describe
both foreshocks, aftershocks and mainshocks.
The abandon of the ingrained concept (in many seismologists' mind)
of the distinction between foreshocks, aftershocks and mainshocks
is an important step towards a simplification and towards an
understanding of the mechanism underlying earthquake sequences.
Ultimately, this parsimonious assumption will be validated or falsified
by the comparison of its prediction with empirical data.
In particular, the deviations from the predictions derived from this
assumption will provide guidelines to enrich the physics.

In order to avoid problems arising from divergences associated with the
proliferation of small earthquakes, the ETAS model assumes the existence of
a magnitude cut-off $m_0$, or equivalently an energy cut-off $E_0$,
such that only earthquakes of magnitude $m \geq m_0$ are allowed to
give birth to aftershocks larger than $m_0$, while events of smaller magnitudes are lost
for the epidemic dynamics. We refer to [{\it Helmstetter and Sornette}, 2002a]
for a justification of this hypothesis and a discussion of ways to improve this
description.

The ETAS model assumes that the seismicity rate (or ``bare Omori propagator'')
at a time between $t$ and $t+dt$, resulting in direct ``lineage'' (without
intermediate events) from an earthquake $i$ that occurred at time
$t_i$, is given by
\be
\phi_{E_i}(t-t_i) = \rho(E_i)~\Psi(t-t_i)~,
\label{first}
\ee
where $\Psi(t)$ is the normalized waiting time density distribution
(that we shall take later given by (\ref{mbjosl}) and
$\rho(E_i)$  defined by
\be
\rho(E_i) =k ~(E_i/E_0)^a
\label{mhmjlr}
\ee
gives the average number of daughters born
from a mother with energy $E_i \geq E_0$.
This term $\rho(E_i)$ accounts for the fact that large
mothers have many more daughters than small mothers because the larger
spatial extension of their rupture triggers a larger domain.
Expression (\ref{mhmjlr}) results in a natural way from the
assumption that aftershocks are events influenced by stress transfer mechanisms
extending over a space domain proportional to the size of
the mainshock rupture [{\it Helmstetter}, 2003].
Indeed, using the well-established scaling law relating the size of rupture and
the domain extension of aftershocks [{\it Kagan}, 2002]
to the release energy (or seismic moment), and assuming a uniform
spatial distribution of aftershocks in their domain,
expression (\ref{mhmjlr}) immediately follows (it still holds
if the density of aftershocks is slowly varying or power law decaying
with the distance from the mainshock).

The value of the exponent $a$ controls the nature of
the seismic activity, that is, the relative role of small compared
to large earthquakes. Few studies have measured $a$
in seismicity data [{\it Yamanaka and Shimazaki}, 1990; {\it Guo and
Ogata}, 1997;
{\it Helmstetter}, 2003].
This parameter $a$ is often found close to the $\beta$ exponent
of the energy distribution defined below in equation (\ref{GRlaw})
[e.g., {\it Yamanaka and K. Shimazaki}, 1990] or fixed arbitrarily equal to
$\beta$ [e.g., {\it Kagan and Knopoff}, 1987; {\it Reasenberg and Jones}, 1989;
{\it Felzer et al.}, 2002]. For a large range of mainshock magnitudes
and using a more sophisticated scaling approach,
{\it Helmstetter} [2003] found $a=0.8\beta$ for the Southern
California seismicity.
If $a < \beta$, small earthquakes, taken together, trigger more
aftershocks than larger
earthquakes. In contrast, large earthquakes dominate earthquake
triggering if $a \geq \beta$.
This case $a \geq \beta$ has been studied analytically in the
framework of the ETAS model
by {\it Sornette and Helmstetter} [2002]
and has been shown to eventually lead to a finite time singularity of
the seismicity rate. This
explosive regime cannot however describe a stationary seismic activity.
In this paper, we will therefore consider only the case $a<\beta$.

An additional space-dependence can be added to $\phi_{E_i}(t-t_i)$
[{\it Helmstetter and Sornette}, 2002b]:
when integrated over all space, the prediction of the space-time model
retrieves those of the pure time-dependent model. Since we are interested
in the inverse Omori law for foreshocks, which is a statement describing
only the time-dependence, it is sufficient to use the time-only version
of the ETAS model for the theory.

The model is complemented by the Gutenberg-Richter law which states that
each aftershock $i$  has an energy $E_i \geq E_0$ chosen according to the
density distribution
\be
P(E) = {\beta E_0^{\beta} \over E^{1+\beta}},  ~~~~{\rm with}~~~\beta  \simeq
2/3~.
\label{GRlaw}
\ee
$P(E)$ is normalized $\int_{E_0}^{\infty} dE ~P(E) =1$.

In view of the empirical observations that the observed rate of
aftershocks decays as a power law of the time since the mainshock, it
is natural to choose the ``bare'' modified Omori law
(or the normalized waiting time distribution between events)
$\Psi(t-t_i)$ in (\ref{first}) also as a power law
\be
\Psi(t-t_i) = {\theta~c^{\theta}  \over (t-t_i+c)^{1+\theta}}~.
\label{mbjosl}
\ee
$\Psi(t-t_i)$ is the rate of daughters of the first generation born at
time $t-t_i$ from the mother-mainshock. Here, $c$ provides
an ``ultra-violet'' cut-off which ensures the
finiteness of the number of aftershocks at early times. It is important
to recognize that the observed aftershock decay rate may be different
from $\Psi(t-t_i)$ due to the effect of aftershocks of aftershocks, and
so on [{\it Sornette and Sornette}, 1999; {\it Helmstetter and
Sornette}, 2002a]

The ETAS model is a ``branching'' point-process
[{\it Harris}, 1963; {\it Daley and Vere-Jones}, 1988]
controlled by the key parameter $n$ defined as the average number (or
``branching ratio'') of daughter-earthquakes
created per mother-event, summed over all times and averaged over all
possible energies.
This branching ratio $n$ is defined as the following  integral
\be
n \equiv \int \limits_0^{\infty} dt 
\int \limits_{E_0}^{\infty} dE~P(E)~\phi_{E}(t)~. 
\label{nvaqlue1}
\ee
The double integral in (\ref{nvaqlue1}) converges if $\theta>0$
and $a<\beta$. In this case, $n$ has a finite value
\be
n={ k \beta \over \beta-a}~,
\label{nvaqlue}
\ee
obtained by using the separability of the two integrals in  (\ref{nvaqlue1}).
The normal regime corresponds to the subcritical case $n<1$ for which
the seismicity rate decays after a mainshock to a constant background (in
the case of a steady-state source) decorated by fluctuations in the seismic
rate.

The total rate of seismicity $\lambda(t)$ at time $t$ is given by
\be
\lambda(t)= s(t) + \sum_{i~|~t_i\leq t} \phi_{E_i}(t-t_i)
\label{mgmkr}
\ee
where $\phi_{E_i}(t-t_i)$ is defined by (\ref{first}).
The sum $\sum_{i~|~t_i\leq t}$ is performed over all events that
occurred at time $t_i\leq t$,
where $E_i$ is the energy of the earthquake that occurred at $t_i$.
$s(t)$ is often taken as a stationary Poisson background stemming from 
plate tectonics and provides a driving source to the process.
The second term in the right-hand-side of expression (\ref{mgmkr}) is
nothing but the sum of (\ref{first}) over all events preceding time $t$.

Note that there are three sources of stochasticity underlying the dynamics
of $\lambda(t)$: (i) the source term $s(t)$ often taken as Poissonian,
(ii) the random occurrences of preceding earthquakes defining the time sequence
$\{t_i\}$ and (iii) the draw of the energy of each event according to the
distribution $P(E)$ given by (\ref{GRlaw}).
Knowing the seismic rate $\lambda(t)$ at time $t$,
the time of the following event is then determined according to a
non-stationary Poisson process of conditional intensity $\lambda(t)$,
and its magnitude is chosen according to the Gutenberg-Richter
distribution (\ref{GRlaw}).

\subsection{The Master equation for the average seismicity rate}

It is useful to rewrite expression (\ref{mgmkr}) formally as
$$
\lambda(t)= s(t) +
$$
\be
\int \limits_{-\infty}^t d\tau~\int \limits_{E_0}^{+\infty} dE~
\phi_{E}(t-\tau)  \sum_{i~|~t_i\leq t} \delta(E-E_{i})~
\delta(\tau-t_i)~,
\label{maagmkr}
\ee
where $\delta(u)$ is the Dirac distribution.
Taking the expectation of (\ref{maagmkr}) over
all possible statistical scenarios (so-called ensemble average),
and assuming the separability in time and magnitude, 
we obtain the following Master equation for
the first moment or statistical average $N(t)$ of $\lambda(t)$
[{\it Helmstetter and Sornette}, 2002a]
\be
N(t) = \mu + \int \limits_{-\infty}^t d\tau~\phi(t-\tau)~N(\tau) ~,
\label{thirdter}
\ee
where $\mu$ is the expectation of the source term $s(t)$ and
\be
\phi(t) \equiv \int \limits_{E_0}^{\infty} dE'~P(E')~ \phi_{E'}(t)~.
\label{mngmldls}
\ee
By virtue of (\ref{nvaqlue}), $\int_0^{\infty} \phi(t) dt =n$.
We have used the definitions
\be
N(t) = \langle \lambda(t) \rangle =  \langle  \sum_{t_i\leq t}
\delta(t-t_i) \rangle~,
\ee
and
\be
P(E)= \langle \delta(E-E_{i}) \rangle~,
\label{mgnnnmv}
\ee
where the brackets $\langle . \rangle$ denotes the ensemble average.
The average is performed
over different statistical responses to the same source term
$s(t)$, where $s(t)$ can be arbitrary.
$N(t) dt $ is the average number of
events occurring between $t$ and $t+dt$ of any
possible energy.

The essential approximation used to derive (\ref{thirdter}) is that
\be
\langle \rho(E_i) \delta(E-E_{i})~\delta(\tau-t_i) \rangle =
\langle \rho(E_i) \delta(E-E_{i}) \rangle ~\langle \delta(\tau-t_i) \rangle
\label{nbbndklk}
\ee
in (\ref{maagmkr}). In words, the fluctuations of the earthquake energies
can be considered to be decoupled from those of the seismic rate.
This approximation is
valid for $a < \beta/2$, for which the random variable $\rho(E_i)$ has
a finite variance.
In this case, any coupling between the fluctuations of the earthquake energies
and the instantaneous seismic rate provides only sub-dominant corrections to the
equation (\ref{thirdter}). For $a > \beta/2$, the variance of $\rho(E_i)$ is
mathematically infinite or undefined as $\rho(E_i)$ is distributed according
to a power law with exponent $\beta/a <2$ (see chapter 4.4 of [{\it
Sornette}, 2000]).
In this case, the Master equation  (\ref{thirdter}) is not completely correct
as an additional term must be included to account for the dominating effect of
the dependence between the fluctuations of earthquake energies and
the instantaneous
seismic rate.

Equation (\ref{thirdter}) is a linear self-consistent
integral equation. In the presence of a stationary source of average
level $\mu$, the average seismicity in the sub-critical
regime is therefore
\be
\langle N \rangle = {\mu \over 1-n}~.
\label{averams}
\ee
This result (\ref{averams}) shows that the effect of the cascade of
aftershocks of aftershocks and so on is to renormalize the average
background seismicity $\langle s \rangle$ to a significantly higher level,
the closer $n$ is to the critical value $1$.

In order to solve for $N(t)$ in the general case, it is convenient to
introduce the Green function or ``dressed propagator'' $K(t)$ defined as
the solution of (\ref{thirdter}) for the case where the source term
is a delta function centered at the origin of time corresponding to a single mainshock:
\be
K(t) = \delta(t) + \int \limits_{0}^t d\tau~\phi(t-\tau)~K(\tau) ~.
\label{thissrdter}
\ee
Physically, $K(t)$ is nothing but the ``renormalized'' Omori law quantifying
the fact that the event at time 0 started a sequence of aftershocks
which can themselves trigger secondary aftershocks and so on.
The cumulative effect of all the possible branching paths
of activity gives rise to the net seismic activity $K(t)$ triggered by
the initial event at $t=0$. Thus, the decay rate of aftershocks
following a mainshock
recorded in a given earthquake catalog is
described by $K(t)$, while $\Psi(t)$ defined by (\ref{mbjosl}) is
a priori unobservable (see however [{\it Helmstetter and Sornette}, 2002a]).

This remark is important because it turns out that the renormalized
Omori law $K(t)$ may be very different from the bare
Omori law $\Psi(t-t_i)$, because of the
effect of the cascade of secondary, tertiary, ..., events triggered
by any single event.
The behavior of the average renormalized Omori law $K(t)$
has been fully classified in [{\it Helmstetter and Sornette}, 2002a]
(see also [{\it Sornette and Sornette}, 1999]):
with a single value of the exponent $1+\theta$ of the ``bare'' propagator
$\Psi(t) \sim 1/t^{1+\theta}$ defined in (\ref{mbjosl}), one obtains a
continuum of apparent exponents for the global rate of aftershocks.
This result may account for the observed variability of
Omori exponent $p$ in the range $0.5-1.5$ or beyond, as
reported by many workers [{\it Utsu et al.}, 1995]. Indeed, the 
general solution of
(\ref{thissrdter}) in the subcritical regime $n<1$ is
\ba
K(t) &\sim& 1/t^{1-\theta}~,~{\rm for}~~c<t<t^*~, \nonumber \\
K(t) &\sim& 1/t^{1+\theta}~,~{\rm for}~~t>t^*~,
\label{soklfm}
\ea
where
\be
t^* \approx c (1-n)^{-1/\theta}~.
\label{formfhs}
\ee
Thus, in practice, the apparent $p$ exponent can be found
anywhere between $1-\theta$ and $1+\theta$.
This behavior (\ref{soklfm})
is valid for $a < \beta/2$ for which, as we explained already, the fluctuations
of the earthquake energies can be considered to be decoupled from those
of the seismic rate.

In the case $a>\beta/2$, this approximation is no more valid
and the problem is considerably more difficult due to the coupling between
the fluctuations in the sequence of earthquake energies and the seismic rate.
We have not been able to derive the detailed solution of the
problem in this regime but nevertheless can predict that the apparent
exponent for the
dressed propagator $K(t)$ should change continuously from $1-\theta$ to
$1+\theta$ as $a$ increases towards $\beta$ from below. The
argument goes as follows. Starting from (\ref{maagmkr}), it is clear that
the larger $a$ is, the larger is the dependence between
the times of occurrences contributing to the sum over
$\delta(\tau-t_i)$ and the
realizations of corresponding earthquake energies contributing
to the sum over $\delta(E-E_{i})$. This is due to the fact that
very large earthquakes trigger many more aftershocks for large $a$, whose
energies influence subsequently the time of occurrences of following
earthquakes,
and so on. The larger is the number of triggered events per shock, the more
intrically intertwined are the times of occurrence and energies of
subsequent earthquakes. 

We have not been able
to derive a full and rigorous analytical treatment of this dependence, yet.
Nevertheless, it is possible to predict the major effect of
this dependence by the following argument.
Consider two random variables $X$ and $Y$, which are (linearly) correlated.
Such a linear correlation is equivalent to the
existence of a linear regression of one variable with respect to the other:
$Y=\gamma X + x$, where $\gamma$ is non-random and is simply related to
the correlation coefficient between $X$ and $Y$ and $x$ is an 
idiosyncratic noise uncorrelated with $X$. Then,
\be
\langle XY \rangle =
\gamma \langle X^2 \rangle + \langle X \rangle ~\langle x \rangle~,
\label{mgmlsa}
\ee
which means that the covariance of $X$ and $Y$ contains a term
proportional to the variance of $X$.

Let us now apply this simple model to the effect of the dependence
between $X \equiv \delta(\tau-t_i)$ and $Y \equiv \rho(E_i) \delta(E-E_{i})$
in the earthquake cascade process. We propose to take 
into account this dependence by the
following ansatz, which corrects (\ref{nbbndklk}), based on 
a description capturing the dependence through the second-order
moment, that is, their covariance:
$$
\langle \rho(E_i) \delta(E-E_{i})~\delta(\tau-t_i) \rangle \approx
$$
\be
\langle \rho(E_i) \delta(E-E_{i}) \rangle ~\langle \delta(\tau-t_i) \rangle
+ \gamma(a) \langle [\delta(\tau-t_i)]^2 \rangle~,
\label{nbbndklaak}
\ee
where $\gamma(a) = 0$ for $a<\beta/2$ and increases with $a>\beta/2$. The
quadratic term
just expresses the dependence between $\rho(E_i) \delta(E-E_{i})$
and $\delta(\tau-t_i)$, i.e., $\rho(E_i) \delta(E-E_{i})$ has a
contribution proportional to $\delta(\tau-t_i)$
as in (\ref{mgmlsa}):  the mechanism leading to the
quadratic term $\langle X^2 \rangle$ is at the source of
$[\delta(\tau-t_i)]^2$ in (\ref{nbbndklaak}).
This new contribution leads
to a modification of (\ref{thissrdter}) according to
\be
K(t) \sim  \int \limits_{0}^t d\tau~\phi(t-\tau)~K(\tau) +
\gamma(a) \int \limits_{0}^t d\tau~\phi(t-\tau)~[K(\tau)]^2~.
\label{thirdeqrraaa}
\ee
Dropping the second term in the right-hand-side of
(\ref{thirdeqrraaa}) recovers
(\ref{thissrdter}). Dropping the first term in the right-hand-side of
(\ref{thirdeqrraaa})
yields the announced result $K(t) \propto 1/t^{1+\theta}$ even in the
regime $t<t^*$.
We should thus expect a cross-over from $K(t) \propto 1/t^{1-\theta}$
to $K(t) \propto 1/t^{1+\theta}$ as $a$ increases from $\beta/2$ to $\beta$.
This prediction is verified accurately
by our numerical simulations.

Once we know the full (ensemble average) seismic response $K(t)$ from
a single event, the complete
solution of (\ref{thirdter}) for the average seismic rate $N(t)$
under the action of the general source term $s(t)$ is
\be
N(t) = \int \limits_{-\infty}^t d\tau~s(\tau) ~K(t-\tau)  ~.
\label{mhjjlss}
\ee
Expression (\ref{mhjjlss}) is nothing but the theorem of Green functions
for linear equations with source terms [{\it Morse and Feshbach}, 1953].
Expression (\ref{mhjjlss}) reflects the intuitive fact that the total
seismic activity at time $t$ is the sum of the contributions of all the
external sources at all earlier times $\tau$ which convey their influence
up to time $t$ via the ``dressed propagator'' (or renormalized Omori law)
$K(t-\tau)$. $K(t-\tau)$ is the relevant kernel quantifying the influence
of each source $s(\tau)$ because it takes into account all possible paths
of seismicity from $\tau$ to $t$ triggered by each specific source.

\subsection{Deviations from the average seismicity rate}

Similarly to the definition (\ref{thissrdter}) of the average
renormalized propagator $K(t)$,
let us introduce the stochastic propagator $\kappa(t)$, defined as
the solution of (\ref{mgmkr}) or (\ref{maagmkr}) for the source term
$s(t)=\delta(t)$. The propagator $\kappa(t)$
is thus the seismicity rate initiated by a single earthquake at the
origin of times, which takes into account the specific sequence of 
generated earthquakes. Since the earthquakes are generated according 
to a probabilistic (generalized Poisson) process, repeating the
history leads in general to different realizations. $\kappa(t)$
is thus fundamentally realization-specific and there are as many different
$\kappa(t)$'s as there are different earthquake sequences. In other words, 
$\kappa(t)$ is a stochastic function.
Obviously,  $\langle \kappa(t) \rangle \equiv K(t)$, that is, its
ensemble average retrieves the average renormalized propagator.

      From the structure of (\ref{mgmkr}) or (\ref{maagmkr}) which
are linear sums over events, an expression similar to (\ref{mhjjlss})
can be written for the non-average seismic rate with an arbitrary 
source term $s(t)$:
\be
\lambda(t) = \int \limits_{-\infty}^t d\tau ~s(\tau)~
\kappa_{\{\tau\}}(t-\tau)~,
\label{mbnn}
\ee
where the subscript $\{\tau\}$ in the stochastic kernel
$\kappa_{\{\tau\}}(t-\tau)$
captures the fact that there is a different stochastic realization of
$\kappa$ for each successive source.
Taking the ensemble average of (\ref{mbnn}) recovers (\ref{mhjjlss}).
The difference between the stochastic kernel $\kappa_{\{\tau\}}(t-\tau)$,
the local propagator $\phi_E(\tau)$ and the renormalized propagator
$K(\tau)$ is illustrated on Figure \ref{fig1} for a numerical simulation
of the ETAS model.

We show in the Appendix A that $\lambda(t)$ can be expressed as
\be
\lambda(t) = N(t) + \int \limits_{-\infty}^t d\tau ~\eta(\tau) ~K(t-\tau)~,
\label{mgktsx}
\ee
where $\eta(\tau)$ is a stationary noise which can be suitably
defined. This is the case because the fluctuations
$\delta P(E)$ of the Gutenberg-Richter law
and of the source $s(t)$ are stationary processes, and because
the fluctuations of $\delta \kappa$ are proportional to $K(t)$.
The expression of $\eta(\tau)$ can be determined explicitly in the case
where the fluctuations of the energy distribution $P(E)$ dominate the
fluctuations
of the seismicity rate $\kappa (\tau)$ (see Appendix A).

\section{Derivation of the inverse Omori law and consequences}

\subsection{Synthesis of the results}

The normal regime in the ETAS model corresponds to the subcritical
case $n<1$ for which
the seismicity rate decays on average after a mainshock to a constant
background (in the case of a steady-state source) decorated by
fluctuations. How is it then possible in this framework to get an
accelerating seismicity preceding a large event?
Conceptually, the answer lies in the fact that when one
defines a mainshock and its foreshocks, one introduces automatically
a conditioning (in the
sense of the theory of probability) in the earthquake statistics.
As we shall see, this conditioning means that specific precursors
and aftershocks must precede and follow a large event.
In other words, conditioned on the observation of a large event, the
sequence of events preceding it cannot be arbitrary.
We show below
that it in fact follows the inverse Omori law in an average
statistical sense.
Figure \ref{fig2} presents typical realizations of
foreshock and aftershock sequences in the ETAS model
as well as the direct
and inverse Omori law evaluated by averaging over many realizations.
The deceleration of the aftershock activity is clearly observed
for each individual sequence as well as in their average. Going to
backward time to compare with
foreshocks, the acceleration of aftershock seismicity when approaching
the main event is clearly visible for each sequence. In contrast,
the acceleration of foreshock activity (in forward time)
is only observable for the ensemble average
while each realization exhibits large fluctuations with no clearly visible
acceleration. This stresses the fact that the inverse Omori law is
a statistical statement, which has a low probability to be observed
in any specific sequence.

Intuitively, it is clear that within the ETAS model, 
an event is more likely to occur after an increase both in seismicity
rate and in magnitudes of the earthquakes, so that this increase of
seismicity can trigger an event with a non-negligible probability.
Indeed, within the ETAS model, all events are the result
of the sum of the background seismicity (due to tectonic forces) and
of all other earthquakes that can trigger their aftershocks.

How does the condition that an earthquake
sequence ends at a mainshock impact on the seismicity prior to that
mainshock? How does this condition create the inverse Omori law?
Since earthquake magnitudes are
independently drawn from the Gutenberg-Richter law, the statistical
qualification of a mainshock, that we place without loss of
generality at the origin of time, corresponds to imposing an anomalous 
burst of seismic activity $\lambda(0)=\langle N \rangle + \lambda_0$ 
at $t=0$ above its average level $\langle N \rangle$
given by (\ref{averams}).
For the study of type II foreshocks (as defined in the introduction), 
we do not constrain the mainshock to be larger than the seismicity
before and after this mainshock. For large mainshock magnitudes,
relaxing this hypothesis will not change the results derived below.

The question then translates into what is the path taken by the noise
$\eta(\tau)$ in (\ref{mgktsx}) for $-\infty < \tau <0$
that may give rise to this burst $\lambda_0$ of activity. The solution is
obtained from the key concept that the set of $\eta(\tau)$'s for $-\infty <
\tau <0$
is biased by the existence of the conditioning, i.e., by the large
value of
$\lambda(0)= \langle N \rangle +\lambda_0$ at $t=0$. This does not mean
that there is an unconditional bias. Rather, the existence of a mainshock
requires
that a specific sequence of noise realizations must have taken place to
ensure its existence.
This idea is similar to the well-known result that an unbiased random
walk $W(t)$ with
unconditional Gaussian increments with zero means sees its position
take a non-zero expectation
\be
\langle W(\tau) \rangle|_c = [W(t)-W(0)]~{\tau \over t}~,
\label{mgnled}
\ee
if one knows the beginning $W(0)$ and the
end $W(t)$ position of the random walk, while the unconditional
expectation $\langle W(\tau) \rangle$ is identically zero. Similarly,
the conditional increment from $\tau$ to $\tau + d\tau$
of this random walk become not non-zero
and equal to (in non-rigorous notation)
\be
d\tau ~{W(t)-W(0) \over t}~,
\label{ngwvckl}
\ee
in contrast with the zero value of the unconditional increments.

In the ETAS model which is a marked point process, the main source of
the noise on $\lambda(t)$ is coming from the ``marks'', that is, the
energies drawn for each earthquake from the Gutenberg-Richter power
law distribution (\ref{GRlaw}). Expression (\ref{mhmjlr}) shows that
the amplitude $\eta_{\tau}$ of the fluctuations in the seismic rate
is proportional to $E_{\tau}^a$, where $E_{\tau}$ is the energy of a
mother-earthquake occurring at time $\tau$.
Since the energies are distributed according to the power law (\ref{GRlaw})
with exponent $\beta$, $\eta_{\tau} \propto E_{\tau}^a$ is distributed
according to a power law with exponent $m=\beta/a$ (see
for instance chapter 4.4 of [{\it Sornette}, 2000]).

We first study the subcritical regime $n<1$ for times $t_c-t<t^*$,
where $t^*$ is defined by (\ref{formfhs}).
Two cases must then be considered.
\begin{itemize}
\item For $a<\beta/2$, $m>2$, the variance and covariance of the noise
$\eta_{\tau}$ exist and one can use conditional covariances to calculate
conditional expectations. We show below that the inverse Omori law
takes the form
\be
{\rm E}[\lambda(t)|\lambda_0] \propto  {\lambda_0 \over (t_c-t)^{1-2\theta}}~,
\label{ids1}
\ee
that is, $p'=1-2\theta$.

\item for $a \geq \beta/2$, $m = \beta/a \leq 2$ and the variance and
covariance
of $\eta_{\tau}$ do not exist: one needs a special treatment based
on stable distributions. In this case, neglecting the coupling between
the fluctuations in the earthquake energies and the seismic rate,
we find that the inverse Omori law
takes the form
\be
{\rm E}[\lambda(t)|\lambda_0] \propto  {\lambda_0 \over (t_c-t)^{1-m\theta}}~.
\label{ids2}
\ee
Taking into account the dependence between the fluctuations in the earthquake
energies and the seismic rate, the exponent $p'$ progressively
increases from $1-2\theta$
towards the value $1+\theta$ of the bare propagator as $a$ goes from
$\beta/2$ to
$\beta$ (see Figure \ref{fig5}).
The increase of $p'$ is thus faster than the dependence $1-m\theta$
predicted by (\ref{ids2}).
\end{itemize}
In the large times limit $t_c-t>t^*$ (far from the
mainshock) of the subcritical regime, we also obtain
an inverse Omori law which takes the form
\be
{\rm E}[\lambda(t)|\lambda_0] \propto  {\lambda_0 \over (t_c-t)^{1+\theta}}~,
~~~~~{\rm for}~~a<\beta/2
\label{ids3}
\ee
and
\be
{\rm E}[\lambda(t)|\lambda_0] \propto  {\lambda_0 \over
(t_c-t)^{1+(m-1)\theta}}~,
~~~~~{\rm for}~~\beta/2 \leq a \leq \beta~.
\label{idsjgjr}
\ee

The direct and inverse Omori laws are clearly observed
in numerical simulations of the ETAS model, when stacking many sequences
of foreshocks and aftershocks, for various mainshock magnitudes
(Figures \ref{fig3} and \ref{fig4}).
Our main result shown in Figure \ref{fig3}
is that, due to conditioning, the inverse Omori law
is {\it different} from the direct Omori law, in that the exponent $p'$
of the inverse Omori law is in general smaller than the exponent
$p$ of the direct Omori law.
Another fundamental difference between aftershocks
and foreshocks found in the ETAS model
is that the number of aftershocks increases as a power $E^a$
of the mainshock energy $E$ as given by (\ref{mhmjlr}),
whereas the number of foreshocks of type II is
independent of the mainshock energy (see Figures \ref{fig3} and \ref{fig4}).
Because in the ETAS model the magnitude of each event is independent of
the magnitude of the triggering events, and of the
previous seismicity, the rate of seismicity increases on average according
to the inverse Omori law before any earthquake, whatever its magnitude.
The number of foreshocks of type I increases with
the mainshock magnitude, for small
and intermediate mainshock magnitudes and saturates to the level of
foreshocks of type II for large mainshocks
because the selection/condition
acting of those defined foreshocks becomes less and
less severe as the magnitude of the mainshock increases
(see Figure \ref{typeI}).
The conditioning that foreshocks of type I must be smaller than their
mainshock induces an apparent increase of the Omori exponent $p'$ as
the mainshock magnitude decreases.
The predictions (\ref{soklfm}) and (\ref{ids1}) on the $p$ and $p'$-value
of type II foreshocks are well-verified by numerical simulations of the
ETAS model up to $a/\beta = 0.5$, as
presented on Figure \ref{fig5}.
However, for $a/\beta > 0.5$,
both $p$ and $p'$ are found larger than predicted by (\ref{soklfm})
and (\ref{ids2}) respectively, due to the coupling between the
fluctuations in the earthquake
energies and those of the seismic rate. This coupling occurs because the
variance of the number $\rho(E)$ of direct aftershocks of an earthquake of
energy $E$ is unbounded for $a> \beta/2$, leading to strong burst of seismic
activity coupled with strong fluctuations of the earthquake energies.
In this regime, expression (\ref{thirdeqrraaa}) shows that
$p$ changes continuously between $1-\theta$ for $a/\beta=0.5$ to
$1+\theta$ for $a=\beta$
in good agreement with the results of the numerical simulations.
In this case $a \geq \beta/2$, the exponent $p'$ is also observed to
increase between
$p'=1-2\theta$ for $a = \beta/2$ to $p'=1+\theta$ for $a=\beta$, as
predicted below.

The dissymetry between the inverse Omori law for foreshocks and the direct
Omori law (\ref{soklfm}) for aftershocks stems from the fact that,
for foreshocks, one observes a seismic rate conditional on a large rate at
the time $t_c$ of the mainshock while, for the aftershocks,
one observes the direct response $K(t)$ to a single large shock.
The later effect stems from the term $\rho(E)$ given by (\ref{mhmjlr})
in the bare Omori propagator which ensures that a mainshock with a
large magnitude triggers aftershocks which dominates overwhelmingly the seismic
activity.
In the special case where one take the exponent $a=0$ in (\ref{mhmjlr}),
a mainshock of large magnitude has no more daughters than any other
earthquake. As a consequence, the observed Omori law stems from the same
mechanism as for the foreshock and the increasing foreshock activity
(\ref{ids1}) gives the same parametric form for the aftershock decay,
with $t_c-t$ replaced by $t-t_c$ (this is for instance
obtained through the Laplace transform of the seismic rate).
This gives the exponent $p=p'=1-2\theta$ for $a=0$ as
for the foreshocks, but the number of aftershocks is still larger than
the number of foreshocks. This result is born out by our numerical
simulations (not shown).

These results and the derivations of the inverse Omori law make clear
that mainshocks are more than just the aftershocks of their foreshocks, as
sometimes suggested [{\it Shaw}, 1993; {\it Jones et al.}, 1999].
The key concept is that all earthquakes are preceded by some seismic
activity and may be seen as the result of this seismic activity.
However, on average, this seismic activity must increase to be compatible
statistically with the occurrence of the main shock: this is an unavoidable
statistical certainty with the ETAS model, that we derive below.
The inverse Omori law is fundamentally a conditional statistical law which
derives from a double renormalization process:
(1) the renormalization from the bare Omori propagator
$\Psi(t)$ defined by (\ref{mbjosl}) into the renormalized or dressed
propagator $K(t)$ and (2) the conditioning of the fluctuations in
seismic activity
upon a large seismic activity associated with the mainshock. In summary,
we can state that mainshocks are aftershocks of conditional foreshocks.
We stress again that the statistical nature of foreshocks does not imply
that there is no information in foreshocks on future large earthquakes.
As discussed below, in the ETAS model, foreshocks are forerunners of 
large shocks.

\subsection{The inverse Omori law $\sim 1/t^{1-2\theta}$ for
$a<\beta/2$}

Let us call $X(t) = \lambda(t)-N(t)$ given by (\ref{mgktsx})
and $Y = \lambda(0)-N(0)$. It is a standard result of
stochastic processes with finite variance and covariance that the
expectation of $X(t)$ conditioned on $Y=\lambda_0$ is given by
[{\it Jacod and Shiryaev}, 1987]
\be
{\rm E}[X(t)|Y=\lambda_0] = \lambda_0~{{\rm Cov}(X(t),Y) \over {\rm
E}[Y^2]}~,
\label{mgmms}
\ee
where ${\rm E}[Y^2]$ denotes the expectation of $Y^2$ and ${\rm
Cov}(X(t),Y)$
is the covariance of $X$ and $Y$. Expression (\ref{mgmms})
recovers the obvious result that ${\rm E}[X(t)|Y=\lambda_0]=0$
if $X$ and $Y$ are uncorrelated.

Using the form (\ref{mgktsx}) for $X(t) = \lambda(t)-N(t)$ and the
fact that $X$ has a finite variance, we obtain
$$
{\rm Cov}(X(t),Y) = \int \limits_{-\infty}^t d\tau  
\int \limits_{-\infty}^0 d\tau'~K(t-\tau)~K(-\tau)
$$
\be{\rm Cov}[\eta(\tau) \eta(\tau')].
\label{kacial}
\ee 
For a dependence structure of $\eta(t)$ falling much faster than the 
kernel $K(t)$, the leading behavior of ${\rm Cov}(X(t),Y)$ is obtained by taking
the limit ${\rm Cov}(\eta(\tau),\eta(\tau'))=\delta(\tau-\tau')$. This yields
\be
{\rm Cov}(X(t),Y) = \int \limits_{-\infty}^t d\tau 
~K(t-\tau)~K(-\tau)~,
\label{mgndfga}
\ee
and
\be
{\rm E}[Y^2] = \int \limits_{-\infty}^0 d\tau ~[K(-\tau)]^2~.
\ee
${\rm E}[Y^2]$ is thus a constant while, for $|t|<t^*$ where $t^*$
is defined in (\ref{formfhs}), ${\rm Cov}(X(t),Y) \sim
1/|t|^{1-2\theta}$.
Generalizing to a mainshock occurring at an arbitrary time $t_c$, this
yields the inverse Omori law
\be
{\rm E}[\lambda(t)|\lambda(t_c)=\langle N \rangle + \lambda_0] =
\langle N \rangle
+ C {\lambda_0 \over (t_c-t)^{1-2\theta}}~,
\label{mgmlks}
\ee
where $C$ is a positive numerical constant.

Expression (\ref{mgmlks}) predicts an inverse Omori law
for foreshocks in the form of an average acceleration of seismicity
proportional to $1/(t_c-t)^{p'}$
with the inverse Omori exponent $p'=1-2\theta$, prior to a mainshock.
This exponent $p'$ is smaller than the exponent $p=1-\theta$ of the
renormalized propagator $K(t)$ describing the direct Omori law for
aftershocks.
This prediction is well-verified by numerical simulations of the ETAS
model shown in Figure \ref{fig3}.

As we pointed out in the introduction, {\it Shaw} [1993] derived the
relationship $p'=2p-1$, which yields $p'=1-2\theta$ for $p=1-\theta$,
based on a clever interestingly incorrect reasoning that we now clarify.
Actually, there are two ways of viewing his argument. The most straightforward
one used by Shaw himself consists in considering
a single aftershock sequence starting at time $0$
from a large mainshock. Let us consider two aftershocks at time
$t-\tau$ and $t$.
Forgetting any constraint on the energies, the earthquake at time $t-\tau$ can
be viewed as a foreshock of the earthquake at time $t$. Summing over
all possible
positions of these two earthquakes at fixed time separation $\tau$ then amounts
to constructing a kind of foreshock count which obeys the equation
\be
\int \limits_0^{+\infty} dt ~K(t-\tau)~K(t)~,
\label{jngbnwakl}
\ee
where $K(t)$ is the number of aftershocks at time $t$. This integral
(\ref{jngbnwakl})
recovers equation (12) of [{\it Shaw}, 1993].
If $K(t) \sim 1/t^p$, this integral predicts a dependence $1/\tau^{2p-1}$
for the effective foreshock activity. This derivation shows that the
prediction $p'=2p-1$ results solely from the counting of pairs
at fixed time intervals in an aftershock sequence. It is a pure product
of the counting process.

We can also view this result from the point of view of the ETAS model. In the
language of the ETAS model, Shaw's formula (12) uses the concept that
a mainshock is an aftershock of a cascade
of aftershocks, themselves deriving from an initial event. This
idea implies that the probability for a mainshock to occur is the sum
over all possible time occurrences of the product of
(i) the probability for an aftershock to be triggered by the initial
event and (ii) the probability that this aftershock triggers its own
aftershock at the time of the mainshock. Shaw uses 
(what corresponds to) the dressed propagator $K(t)$ for the
first probability. He also assumes that the rate
of mainshocks deriving from an aftershock of the initial event
is proportional to $K(t)$.
However, from our previous studies [{\it Sornette and Sornette}, 1999;
{\it Helmstetter and Sornette}, 2002a] and the 
present work, one can see that this corresponds to
an illicit double counting or double renormalization.
This danger of double counting is illustrated by comparing the
formulas (\ref{thirdter}, \ref{thissrdter}) with (\ref{mhjjlss}).
Either the direct tectonic source of seismicity $s(t)$ impacts
the future seismicity by a weight given by the renormalized or dressed propagator
as in (\ref{mhjjlss}). Or we can forget about the tectonic source term $s(t)$,
we only record all past seismic activity (all sources and all triggered events) as in
(\ref{thirdter}, \ref{thissrdter}),
but then the impact of all past seismicity on future seismicity is weighted
by the bare propagator. These two view points are completely equivalent
and are two alternative expressions of the Green theorem.
What is then the reason for the correct $1/t^{1-2\theta}$ 
inverse Omori law derived by {\it Shaw} [1993]?
It turns out that his (erroneous) double counting recovers the mathematical
form resulting from the effect of the
conditioning of past seismicity leading to $s(t) \sim K(t)$
valid for $a \leq \beta/2$ as
derived below in (\ref{nbvkd}). Indeed, inserting $s(t) \sim K(t)$
in (\ref{mhjjlss}) retrieves the correct prediction $1/t^{1-2\theta}$
for the inverse Omori law. This proportionality
$s(t) \sim K(t)$ is physically at the
origin of (\ref{mgndfga}) at the origin itself of the inverse Omori law.
In other words, Shaw obtains the correct result (\ref{jngbnwakl})
by incorrect double counting while the correct way to get (\ref{jngbnwakl}) is
that the mainshock is conditional on a specific average trajectory 
of past seismicity captured by $s(t) \sim K(t)$. 
In addition to provide a more correct reasoning, our approach allows
one to explore the role of different parameter regimes and, in particular, 
to analyze the failure of the argument 
for $a > \beta/2$, as already explained.

\subsection{The inverse Omori law $\sim 1/t^{1-\theta \beta/a}$ for
$a \geq \beta/2$}

Expression (\ref{mgktsx}) defines the fluctuating part $X(t) =
\lambda(t)-N(t)$ of the seismic rate as a sum of random variables
$\eta(\tau)$ with power law distributions weighted by the kernel $K(t-\tau)$.
These random variables $\eta(\tau)$, which are mainly dominated
by the fluctuations in event magnitudes but also receive contributions from the
intermittent seismic rate, are conditioned by the realization
of a large seismicity rate
\be
X(0)=\lambda_0 = \int \limits_{-\infty}^0 d\tau ~\eta(\tau) ~K(-\tau)~,
\label{mnbjsl}
\ee
which is the correct statistical implementation of the condition of the
existence of a large shock at $t=0$. Since the conditioning is
performed on $X(0)$, that is,
upon the full set of noise realizations acting up to time $t=0$, the
corresponding conditional noises up to time $t<0$ contribute all to ${\rm
E}[X(t)|X(0)=\lambda_0]_{t<0}$ by their conditional expectations as
\be
{\rm E}[X(t)|X(0)=\lambda_0]_{t<0}= \int \limits_{-\infty}^t d\tau ~
{\rm E}[\eta(\tau)|X(0)] ~K(t-\tau)~.
\label{mgktsaax}
\ee

In Appendix B, it is shown that, for identically independently distributed
random variables $x_i$ distributed according to a power law with exponent
$m=\beta/a \leq 2$ and entering the sum
\be
S_N=\sum_{i=1}^N K_i x_i
\label{makqjq}
\ee
where the $K_i$ are arbitrary positive weights, the expectation ${\rm
E}[x_i|S_N]$ of $x_i$ conditioned on the existence of a large realization
of $S_N$ is given by
\be
{\rm E}[x_i|S_N] \propto S_N ~K_i^{m-1}~.
\label{nbvkd}
\ee

To apply this result to (\ref{mgktsaax}), it is convenient to discretize it.
Some care should however be exercised in this discretization (1) to account
for the expected power law acceleration of ${\rm E}[X(t)|X(0)=\lambda_0]$
up to $t=0$ and (2) to discretize correctly the random noise. We thus write
$$
\int \limits_{-\infty}^0 d\tau ~\eta(\tau) ~K(-\tau)  \approx
\sum_{\tau_i<0} \int \limits_{\tau_i}^{\tau_{i+1}} d\tau ~\eta(\tau) ~K(-\tau)
$$
\be
\sim \sum_{\tau_i<0} (\tau_{i+1}-\tau_{i})~K(-\tau_i)~x_i~,
\label{nglqa}
\ee
where $x_i \sim \eta_i (\tau_{i+1}-\tau_{i})$ is the stationary discrete noise
distributed according to a power law distribution with exponent $m=\beta/a$.
The factor $(\tau_{i+1}-\tau_{i}) \propto |\tau_{i}|$ in front of the
kernel $K(-\tau_i)$
is needed to regularize the discretization
in the presence of the power law acceleration up to time $0$.
In the notation of Appendix B, $(\tau_{i+1}-\tau_{i})~K(-\tau_i)
\propto |\tau_{i}|~K(-\tau_i) \sim 1/|\tau_i|^{-\theta}$ plays the role
of $K_i$. We also need an additional factor $(\tau_{i+1}-\tau_{i})$ to obtain
a regularized noise term: thus,
$\eta_i (\tau_{i+1}-\tau_{i}) \propto \eta_i |\tau_{i}|$
plays the role of $x_i$. This discretization procedure recovers the
results obtained by using (\ref{mgmms}) and the variance and covariance
of the continuous integrals for the case $a<\beta/2$ where they are defined.
Note that the last expression in equation (\ref{nglqa}) does not keep track
of the dimensions as we are only able to obtain the leading scaling behavior
in the discretization scheme.

Using (\ref{nbvkd}), we thus obtain
${\rm E}[\eta_i|\tau_{i}| ~ |X(0)=\lambda_0] \propto {\lambda_0 \over
|\tau_i|^{-\theta(m-1)}}$
and thus
\be
{\rm E}[\eta_i|X(0)=\lambda_0] \propto {\lambda_0 \over
|\tau_i|^{1-\theta(m-1)}}~.
\label{nfgla}
\ee

Similarly to (\ref{nglqa}), the discrete equivalent to (\ref{mgktsaax})
reads
\ba
&& {\rm E}[X(t)|X(0) = \lambda_0]_{t<0}   \label{mgktsaaaaaax} \\
&& \approx \sum_{\tau_i<t} (\tau_{i+1}-\tau_{i})~K(t-\tau_i)~
{\rm E}[\eta_i|\tau_{i}|~|X(0) = \lambda_0] \nonumber \\
&& \sim \int \limits_{-\infty}^t d\tau~
{1 \over |t-\tau|^{1-\theta}}~{\lambda_0 \over |\tau|^{1-\theta(m-1)}}
\sim {\lambda_0 \over |t|^{1-m\theta}}~, \nonumber
\ea
where we have re-introduced the factors $\tau_{i+1}-\tau_{i}$ to reverse
to the continuous integral formulation and have use the definition $m=\beta/a$.
Expression (\ref{mgktsaaaaaax}) gives the inverse Omori law
\be
{\rm E}[X(t)|X(t_c)=\lambda_0]_{t<0} \propto {\lambda_0 \over
(t_c-t)^{1-\theta \beta/a}}
\label{pmse}
\ee
for foreshock activity prior to a mainshock occurring at time $t_c$.
Note that the border case $m=\beta/a=2$ recovers our previous result
(\ref{mgmlks}) as it should.

The problem is that this derivation does not take into account
the dependence between the fluctuations in the earthquake
energies and the seismic rate, which become prominent precisely
in this regime $\beta/2 \leq a \leq \beta$. We have not been able
yet to fully solve this problem for arbitrary values $a$ but
can nevertheless predict that (\ref{pmse}) must be replaced by
\be
{\rm E}[X(t)|X(0)=\lambda_0]_{t<0} \propto {\lambda_0 \over
|t|^{1+\theta}}~,~~~{\rm for}~~a \to \beta~.
\label{pmsaaae}
\ee
We follow step by step the
reasoning from expression (\ref{mgktsaax}) to (\ref{mgktsaaaaaax}),
with the following modifications imposed by
the regime  $\beta/2 \leq a \leq \beta$.
\begin{enumerate}
\item  The conditional expectations given by (\ref{nbvkd})
must be progressively changed
into ${\rm E}[x_i|S_N] \propto S_N ~K_i$ as $a \to \beta$, due to the
coupling between energy and seismic
rate fluctuations (leading to (\ref{nbbndklaak}) via the mechanism 
(\ref{mgmlsa})).
Indeed, the coupling
between energy and seismic rate fluctuations gives rise to the dependence
${\rm E}[x_i|S_N] \propto K_i$ which becomes dominant over
the conditional expectations given by (\ref{nbvkd}) for $m<2$.

\item As shown with (\ref{thirdeqrraaa}), the
dependence between the fluctuations in the earthquake
energies and the seismic rate leads to change
$K(t) \propto 1/t^{1-\theta}$ into
$K(t) \propto 1/t^{1+\theta}$ as $a \to \beta$
even in the regime $t<t^*$.
\end{enumerate}
This leads finally to
changing expressing (\ref{mgktsaaaaaax}) into
\be
{\rm E}[X(t)|X(0)] \sim \int \limits_{-\infty}^t d\tau~
{1 \over |t-\tau +c|^{1+\theta}}~{\lambda_0 \over |\tau|^{1+\theta}}~,
\label{mgktaasaaaaaax}
\ee
where we have re-introduced the regularization constant $c$ to ensure
convergence for $\tau \to t$. Taking into account the
contribution $\propto t^{\theta}$
at this upper bound $t$ of the integrand $\propto
1/|t-\tau +c|^{1+\theta}$, we finally get (\ref{pmsaaae}).
This result is verified numerically in Figure \ref{fig5}.

\subsection{The inverse Omori law in the regime $t_c-t>t^*$}

The inverse Omori laws derived in the two preceding sections are
valid for $t_c-t<t^*$, that is, sufficiently close to the mainshock.
A similar inverse Omori law is also obtained for $t_c-t>t^*$.
In this goal, we use (\ref{soklfm}) showing that the propagator
$K(t-\tau) \propto 1/(t-\tau)^{1-\theta}$ must be replaced by
$K(t-\tau) \propto 1/(t-\tau)^{1+\theta}$ for time difference larger
than $t^*$.
It would however be incorrect to deduce that we just have to
change $-\theta$ into $+\theta$ in expressions (\ref{mgmlks})
and (\ref{pmse}), because the integrals leading to these
results behave differently: as in (\ref{mgktaasaaaaaax}), one
has to re-introduced the regularization constant $c$ to ensure
convergence for $\tau \to t$ of $1/|t-\tau +c|^{1+\theta}$. The
final results are thus given by (\ref{mgmlks})
and (\ref{pmse}) by changing $-\theta$ into $+\theta$ and by
multiplying these expressions by the factor $t^{\theta}$ stemming
from the regularization $c$. Thus,
in the large time limit $t_c-t>t^*$ (far from the
mainshock) of the subcritical regime, we also obtain
an inverse Omori law which takes the form (\ref{ids3})
for $a<\beta/2$ and the form (\ref{idsjgjr}) for $\beta/2 \leq a \leq \beta$.
These predictions are in good agreement with our numerical simulations.

\section{Prediction for the Gutenberg-Richter distribution of foreshocks}

We have just shown that the stochastic component of the seismic rate
can be formulated as a sum of the form (\ref{makqjq}) of variables $x_i$
distributed according to a power law with exponent $m=\beta/a$ and
weight $K_i$.
It is possible to go beyond the derivation of the conditional expectation
${\rm E}[x_i|S_N]$ given by (\ref{nbvkd}) and obtain the conditional
distribution
$p(x_i|S_N)$ conditioned on a large value of the realization of $S_N$.

For this, we use the definition of conditional probabilities
\be
p(x_i|S_N) = {p(S_N|x_i) p(x_i) \over P_N(S_N)}~,
\ee
where $P_N(S_N)$ is the probability density function of the sum $S_N$.
Since $p(S_N|x_i)$ is simply given by
\be
p(S_N|x_i) = P_{N-1}(S_N-K_ix_i)~,
\ee
we obtain
\be
p(x_i|S_N) = p(x_i) {P_{N-1}(S_N-K_i x_i)\over P_N(S_N)}~.
\label{mgnlwlq}
\ee
This shows that the conditional Gutenberg-Richter distribution
$p(x_i|S_N)$
is modified by the conditioning according to the multiplicative
correcting
factor $P_{N-1}(S_N-K_i x_i)/P_N(S_N)$. For large $N$, $P_N$ and
$P_{N-1}$ tend
to stable L\'evy distributions with the same index $m$ but different
scale factors
equal respectively to $\sum_j K_j^m$ and $\sum_{j\neq i} K_j^m$. The
tail of
$p(x_i|S_N)$ is thus
\be
p(x_i|S_N) \sim \left(1- {K_i^m \over \sum_j K_j^m}\right) {1 \over
x_i^{1+m}}~
{1 \over \left( 1-(K_i x_i/S_N) \right)^{1+m}}~.
\label{jgnwq}
\ee
Since $K_i x_i \ll S_N$, we can expand the last term in the
right-hand-side of
(\ref{jgnwq}) and obtain
\be
p(x_i|S_N) \sim \left(1- {K_i^m \over \sum_j K_j^m}\right)
\left[ {1 \over x_i^{1+m}} + (1+m) (K_i/S_N) {1 \over x_i^{m}}
\right]~.
\label{jssgnaawq}
\ee
Since $x_i \sim E_i^a$, we use the transformation property
of distribution functions $p(x_i) dX_i = p(E_i) dE_i$ to obtain the
pdf of foreshock energies $E_i$.
Going back to the continuous limit in
which $K_i/S_N \sim (t_c-t)^{-(1-\theta)}/(t_c-t)^{-(1-\beta\theta)}
=1/(t_c-t)^{(\beta - 1)\theta}$, we
obtain the conditional Gutenberg-Richter distribution for foreshocks
\be
P(E|\lambda_0) \sim {E_0^{\beta} \over E^{1+\beta}} +
{C \over (t_c-t)^{\theta(\beta-a)/a}}{E_0^{\beta'} \over E^{1+ \beta'}}~
\label{GRaalaw}
\ee
where
\be
\beta'=\beta-a~,
\label{gnnlala}
\ee
and $C$ is a numerical constant.
The remarkable prediction (\ref{GRaalaw}) with (\ref{gnnlala})
is that the Gutenberg-Richter distribution is modified upon the approach
of a mainshock by developing a bump in its tail. This modification takes
the form of an additive power law contribution with a new ``$b$-value''
renormalized/amplified by the exponent $a$ quantifying the dependence
of the number of daughters as a function of the energy of the mother.
Our prediction is validated very clearly by numerical simulations reported
in Figures \ref{fig6} and \ref{fig7}.


\section{Migration of foreshocks towards the mainshock}

By the same mechanism leading to (\ref{mgmlks}) via (\ref{mgmms}) and
(\ref{mgndfga}), conditioning the foreshock seismicity to culminate
at a mainshock at time $t_c$ at some point $\vec r$ taken as the origin
of space must lead to a migration towards the mainshock.
The seismic rate $\lambda({\vec r},t)$ at position $\vec r$ at time
$t<t_c$ conditioned on the existence of the mainshock at position
$\vec 0$ at time $t_c$ is given by
\be
{\rm E}[\lambda({\vec r},t)|\lambda({\vec 0}, t_c)]
\sim \int \limits_{-\infty}^t d\tau \int d{\vec \rho}  
~K({\vec r}-{\vec \rho}, t-\tau)~ K({\vec \rho}, t_c-\tau)~.
\label{mgmsslks}
\ee
$K({\vec r}-{\vec \rho}, t-\tau)$ is the dressed spatio-temporal
propagator giving the seismic activity at position $\vec r$ and time $t$
resulting from a triggering earthquake that occurred at position ${\vec \rho}$
at a time $\tau$ in the past. Its expression is given in [{\it
Helmstetter and Sornette}, 2002b] in a variety of situations.
Assuming that the probability distribution for an earthquake to trigger
an aftershock at a distance $r$  is of the form
\be
\rho(r) \sim 1/(r+d)^{1+\mu}~,
\label{rhor}
\ee
{\it Helmstetter and Sornette} [2002b] have shown
that the characteristic size of the
aftershock area  slowly diffuses according to $R \sim t^H$,
where the time $t$ is counted from the time of the mainshock.
For simplicity, $d$ is taken independent of the mainshock energy.
$H$ is the Hurst exponent characterizing the diffusion given by
\be
H={\theta \over \mu}~~~ \mbox{for} ~~ \mu<2~,~~~H={\theta
\over2}~~~\mbox{for} ~~ \mu>2~.
\label{H}
\ee
This diffusion is captured by the fact that $K({\vec r}-{\vec \rho},
t-\tau)$ depends on
${\vec r}-{\vec \rho}$ and $t-\tau$ essentially through the reduced
variable $|{\vec r}-{\vec \rho}|/(t-\tau)^H$. Then,
expression (\ref{mgmsslks}) predicts that this diffusion must be reflected
into an inward migration of foreshock seismicity towards the mainshock
with the same exponent $H$.

These results are verified by numerical simulations of the ETAS model.
Figure \ref{fig8} presents the migration of foreshock activity for
two numerical simulations of the ETAS model, with different parameters.
As for the inverse Omori law, we have superposed many sequences of foreshock
activity to observe the migration of foreshocks.
For a numerical simulation with parameters $n=1$, $\theta=0.2$,
   $\mu=1$, $d=1$ km, $c=0.001$ day, $a=0.5\beta$ and $m_0=2$, we see clearly 
the localization
   of the seismicity as the mainshock approaches.
We obtain an effective migration exponent $H=0.18$, describing how the
effective size $R$ of the cloud of foreshocks shrinks as time $t$
approaches the time $t_c$ of the main shock: $R \sim (t_c-t)^{H}$
(see Figure \ref{fig8}a,c).
This result is in good agreement with the prediction $H=0.2$
given by (\ref{H}).
   The spatial distribution of foreshocks around the mainshock is similar
   to the distribution of aftershocks around the mainshock.
Figure \ref{fig8}b,c presents the migration of foreshock activity for
a numerical simulation with $\theta=0.01$, $\mu=1$, $d=1$ km, $c=0.001$ day 
leading to a very small diffusion exponent $H=0.01$.
The analysis of this foreshock sequence gives an effective
migration exponent $H=0.04$ for short times, and a faster
apparent migration at longer times
due to the influence of the background activity. See
[{\it Helmstetter and Sornette}, 2002b] for a discussion
of artifacts leading to apparent diffusions of seismicity resulting
from various cross-over phenomena.


\section{Discussion}

It has been proposed for decades that many large earthquakes
were preceded by an unusually high seismicity rate, for times
of the order of weeks to months before the mainshock
[{\it Omori}, 1908; {\it Richter}, 1958; {\it Mogi}, 1963].
Although there are large fluctuations in the foreshock patterns from
one sequence to another one, some recurrent properties are observed.
\begin{enumerate}
\item[(i)] The rate of foreshocks
increases as $1/(t_c-t)^{p'}$ as a
function of the time to the main shock at $t_c$,
with an exponent $p'$ smaller than or equal to the
exponent $p$ of direct Omori law;
\item[(ii)] the Gutenberg-Richter
distribution of magnitudes is modified as the mainshock approaches, and
is usually modeled by a decrease in $b$-value;
\item[(iii)] The epicenters of the
foreshocks seem to migrate towards the mainshock.
\end{enumerate}
We must acknowledge that the robustness of these three laws
decreases from (i) to (iii).
In previous sections, we have shown that these properties of
foreshocks derive simply
from the two most robust empirical laws of earthquake occurrence,
namely the Gutenberg-Richter and Omori laws, which define the ETAS model.
In this ETAS framework, foreshock sequences emerge on average by conditioning
seismicity to lead to a burst of seismicity at the time of the mainshock.
This analysis differs from two others
analytical studies of the ETAS model [{\it Helmstetter and Sornette}, 2002b;
{\it Sornette and Helmstetter}, 2002], who proposed that accelerating
foreshock sequences may be related either to the super-critical regime
$n>1$ or to the singular regime $a>\beta$ (leading formally to
$n \to \infty$) of the ETAS model.
In these two regimes, an accelerating seismicity sequence arises from the
cascade of aftershocks that trigger on average more than one
aftershock per earthquake.
Here we show that foreshock sequences emerge in the stationary
sub-critical regime ($n<1$) of the ETAS model, when an event triggers on
average less than one aftershock.
In this regime, aftershock have a low probability of triggering a
larger earthquake.
Nonetheless, conditioning on a high seismicity rate at the time of
the mainshock,
we observe, averaging over many mainshocks, an increase of the seismicity rate
following the inverse Omori law. In addition, as we shall show below, this
increase of seismicity has a genuine and significant predictive power.

\subsection{Difference between type I and type II foreshocks}

Our results applies to foreshocks of type II, defined as earthquakes
preceding a mainshock in a space-time window preceding a mainshock,
independently of their magnitude.
This definition is different from the usual definition of foreshocks, which
imposes a mainshock to be larger than the foreshocks (foreshocks of type I
in our terminology).
Using the usual definition of foreshocks in our numerical simulations
of the ETAS model, our results remain robust but there are quantitative
differences introduced by the somewhat arbitrary
constraint entering into the definition of foreshocks of type I:
\begin{enumerate}
\item a roll-off in the inverse Omori-law,
\item a dependence of
the apparent exponent $p'$ on the time window used to define foreshocks
and mainshocks and
\item a dependence of the rate of foreshocks and of $p'$ on the
mainshock magnitude.
\end{enumerate}
As seen in Figure \ref{typeI},
these variations between foreshocks of type I and type II
are  observed only for small mainshocks. Such foreshocks
are less likely the foreshocks of a mainshock and
are more likely to be preceded by a larger earthquake, that is, to be
the aftershocks of a large preceding mainshock. These subtle
distinctions should attract the attention of the reader
on the arbitrariness underlying the definition of
foreshocks of type I and suggest, together with our results, that
foreshocks of type II are more natural objects to define and study
in real catalogs.
This will be reported in a separate presentation.

\subsection{Inverse Omori law}

Conditioned on the fact that a mainshock is associated with a burst
of seismicity,
the inverse Omori law arises from the expected fluctuations of the
seismicity rate
leading to this burst of seismicity.
Depending on the branching ratio $n$ and on the ratio $a/\beta$, the exponent
$p'$ is found to vary between $1-2\theta$
and $1+\theta$, but is always found to
be smaller than the $p$ exponent of the direct Omori law.
Our results thus reproduce both the variability of $p'$
and the lower value measured for $p'$ than for $p$ reported by
[{\it Papazachos}, 1973, 1975b; {\it Page}, 1986;
{\it Kagan and Knopoff}, 1978; {\it Jones and Molnar}, 1979;
{\it Davis and Frohlich}, 1991; {\it Shaw}, 1993;
{\it Utsu et al.}, 1995; {\it Ogata et al.}, 1995; {\it Maeda}, 1999].
In their synthesis of all $p$ and $p'$ values, 
{\it Utsu et al.} [1995] report $p'$-value in the range 0.7-1.3 , while
$p$ of aftershocks ranges from 0.9 to 1.5.
The few studies that have measured simultaneously $p$ and $p'$ using a
superposed epoch analysis have obtained $p'$ either roughly equal to $p$
   [{\it Kagan and Knopoff}, 1978; {\it Shaw}, 1993]
or smaller than $p$ [{\it Davis and Frohlich}, 1991;
{\it Ogata et al.}, 1995; {\it Maeda}, 1999].
The finding that $p \approx p' \approx 1$ suggested by
[{\it Shaw}, 1993; {\it Reasenberg}, 1999] for the California seismicity
can be interpreted in our framework
as either due to a  very small  $\theta$ value, or due to a large
$a/\beta$ ratio close to $0.8$, as shown in Figures \ref{fig4} and \ref{fig5}.
The result $p' < p$ reported by [{\it Maeda}, 1999] for the Japanese
seismicity and by {\it Davis and Frohlich} [1991] for the worldwide
seismicity can be related to a rather small $a/\beta$ ratio,
as also illustrated in Figures \ref{fig3} and \ref{fig5}.

In contrast with the direct Omori law, which is clearly observed after
all large shallow earthquakes, the inverse Omori law is an average
statistical law,
which is observed only when stacking many foreshock sequences.
Simulations reported in Figure \ref{fig2} illustrate that, for individual
foreshock sequences, the inverse Omori law is difficult to capture.
Similarly to what was done
for real data [{\it Kagan and Knopoff}, 1978; {\it Jones and Molnar}, 1979;
{\it Davis and Frohlich}, 1991; {\it Shaw}, 1993; {\it Ogata et al.}, 1995;
{\it Maeda}, 1999;  {\it Reasenberg}, 1999], the inverse Omori law
emerges clearly in our model only when using a
superposed epoch analysis to average the seismicity rate over a large
number of sequences.
Our results are thus fundamentally different from the critical point theory
[{\it Sammis and Sornette}, 2002] which leads to a power-law increase of
seismic activity preceding each single large earthquake over what is probably
a larger space-time domain [{\it Keilis-Borok and Malinovskaya}, 1964;
{\it Bowman et al}., 1998]. The inverse Omori law is indeed usually observed
for time scales of the order of weeks to months before a mainshock, while  
{\it Keilis-Borok and Malinovskaya} [1964] and {\it Bowman et al} [1998] 
report a precursory increase of seismic activity for time scales
of years to decades before large earthquakes. Our results can thus be considered
as providing a null-hypothesis against which to test the critical point theory.

\subsection{Foreshock occurrence rate}

In term of occurrence rate, foreshocks are less frequent than aftershocks
(e.g. [{\it Kagan and Knopoff}, 1976, 1978; {\it Jones and Molnar}, 1979]).
The ratio of foreshock to aftershock numbers is close to 2-4 for
$M=5-7$ mainshocks,
when selecting foreshocks and aftershocks at a distance $R=50-500$ km
from the mainshock
and for a time $T=10-100$ days before or after the mainshock
   [{\it Kagan and Knopoff} 1976; 1978; {\it Jones and Molnar}, 1979;
{\it von Seggern et al.}, 1981;
   {\it Shaw}, 1993].
In our simulations, large mainshocks have significantly more
aftershocks than foreshocks,
in agreement with observations, while small earthquakes have roughly
the same number of foreshocks (of type II) and of aftershocks.
The ratio of aftershocks to foreshock of type II increases if the ratio
$a/\beta$ decreases, as observed when comparing the case $a=0.5\beta$
shown in Figure \ref{fig3} with the results obtained in the case
$a=0.8\beta$ represented in Figure \ref{fig4}.
This may be explained by the relatively larger weights of the largest
earthquakes which increase with increasing $a$, and by our definition
of aftershocks and foreshocks: recall that aftershock sequences are conditioned
on not being preceded by an event larger than the mainshock, whereas
a foreshock of type II can be larger than the mainshock.
Thus, for large $a/\beta<1$, most ``mainshocks'', according
to our definition, are aftershocks of a preceding large earthquake, whereas
aftershock sequences cannot be preceded by an earthquake larger than
the mainshock.

The retrospective foreshock frequency, that is, the fraction of
mainshocks that are
preceded by a foreshock, is reported to range from 10\% to 40\%
using either regional or worldwide catalogs [{\it Jones and
Molnar}, 1979; {\it von Seggern et al.}, 1981; {\it Yamashina}, 1981;
{\it Console et al.}, 1983; {\it Jones}, 1984; {\it Agnew and Jones}, 1991;
{\it Lindh and Lim}, 1995; {\it Abercrombie and Mori}, 1996; {\it
Michael and Jones}, 1998;
{\it Reasenberg}, 1999]. The variability of the foreshock rate is closely
related to the catalog threshold for the magnitude completeness for the
small events [{\it Reasenberg}, 1999]. These results are in line with
our simulations.

The observed number of foreshocks per mainshock
slowly increases with the mainshock magnitude
[e.g. data from {\it Kagan and Knopoff}, 1978; {\it Shaw}, 1993; {\it
Reasenberg}, 1999].
In our model, the number of foreshocks of type II
is independent of the mainshock magnitude,
because the magnitude of each earthquake is independent
of the previous seismicity history. An increase of the number of
foreshocks of type I as a function of the mainshock magnitude is observed
in our numerical simulations (see Figure \ref{typeI})
because, as we explained before, the constraint on the foreshock magnitudes
to be smaller than the mainshock magnitude is less severe for larger
earthquakes
and thus filter out less foreshocks.
Therefore, our results can explain the increase in the foreshock frequency
with the mainshock magnitude reported using foreshocks of type I.
The slow increase of the number of foreshocks with the mainshock magnitude,
  if any, is different from the predictions of both the nucleation
model [{\it Dodge et al.}, 1996] and of the critical point theory
[{\it Sammis and Sornette}, 2002] which predict an increase
of the foreshocks rate and of the foreshock zone with the mainshock size.

\subsection{Magnitude distribution of foreshocks}

Many studies have found that the apparent $b$-value of the
magnitude distribution of foreshocks is smaller than that of
the magnitude distribution of the background seismicity and of
aftershocks.
Case histories analyze individual foreshock sequences, most of
them being chosen a posteriori to suggest
that foreshock patterns observed in acoustic emissions
preceding rupture in the laboratory could
apply to earthquakes [{\it Mogi}, 1963; 1967].
A few statistical tests validate the significance of reported anomalies
on $b$-value of foreshocks.
A few others studies use a stacking  method to average over many sequences
in order to increase the number of events.

A $b$-value anomaly,  usually a change in the mean $b$-value, for earthquakes
preceding a mainshock has been proposed as a possible precursor on many
retrospective case studies [{\it Suyehiro}, 1966; {\it Papazachos et al.}, 1967;
{\it Ikegami}, 1967; {\it Berg}, 1968; {\it Bufe}, 1970; {\it
Fedotov et al.}, 1972;
   {\it  Wyss and Lee}, 1973; {\it Papazachos}, 1975a,b;
{\it Ma}, 1978; {\it Li et al.}, 1978; {\it Wu et al.} 1978; {\it
Cagnetti and Pasquale}, 1979;
   {\it Stephens et al.}, 1980; {\it Smith}, 1981, 1986; {\it Imoto} 1991;
{\it Enescu and Kito}, 2001].
Most case histories argue for a decrease of $b$-value, but this
decrease, if any, is
sometimes preceded by an increase of $b$-value [{\it Ma},
1978; {\it Smith}, 1981, 1986; {\it Imoto} 1991].
In a couple of cases, temporal decreases in $b$-value before Chinese
earthquakes
were used to issue successful predictions [{\it Wu et al.}, 1978;
{\it Zhang et al.}, 1999].

Because of the paucity of the foreshock numbers, most of the study of
individual
sequences does not allow to estimate a robust temporal change of $b$-values
before mainshocks, nor to characterize the shape of the magnitude distribution.
A few studies have demonstrated the statistical significance of decreases
of $b$-value when the time to the mainshock decreases using a
superposed epoch analysis [{\it Kagan and Knopoff}, 1978;
{\it  Molchan and Dmitrieva, 1990}; {\it Molchan et al.}, 1999].
Using 200 foreshocks sequences of regional and worldwide seismicity,
{\it Molchan et al.} [1999] found
that the $b$-value is divided by a factor approximately
equal to $2$ a few days or hours before the mainshock.
{\it Knopoff et al.} [1982] found no significant differences between
the $b$-value of
aftershocks and foreshocks when investigating
12 individual sequences of California catalogs.
When all the aftershocks and foreshocks in a given
catalog are superposed, the same study showed for catalogs of large durations
(e.g. ISC, 1964-1977; NOAA, 1965-1977) that the $b$-value for foreshocks is
significantly smaller than the $b$-value for aftershocks
[{\it Knopoff et al.}, 1982].
The same pattern being simulated by a branching model for seismicity,
{\it Knopoff et al.} [1982]
surmise that the observed and simulated changes in magnitude distribution
value arises intrinsically from the conditioning of aftershocks and foreshocks
and from the smaller numbers of foreshocks relatively to aftershocks numbers
when counted from the mainshock time. The result of
[{\it Knopoff et al.}, 1982] is often cited as disproving the reality of
a change of $b$-value. Our results find that a change in $b$-value in the
ETAS branching model of seismicity is a physical phenomenon with real
precursory content. This shall be stressed further below in association with
Figure \ref{errordiagram}. Therefore, the fact that a change in $b$-value
can be reproduced by a branching model of seismicity cannot discredit
the strong empirical evidence of a change of $b$-value
[{\it Knopoff et al.}, 1982] and its genuine
physical content capturing the interactions between and triggering of
earthquakes.

The observed modification of the magnitude distribution of foreshocks is
usually interpreted as a decrease of $b$-value as the mainshock approaches.
However, some studies argue that the Gutenberg-Richter distribution before
a mainshock is no more a pure power-law distribution, due to an apparent
increase of the number of large events relatively to the Gutenberg-Richter law,
while the rate of small earthquakes remains constant.
Such pattern is suggested by {\it Rotwain et al.} [1997] for both
acoustic emission
preceding material failure, and  possibly for Californian seismicity
preceding large earthquakes.
Analysis of seismicity before recent large shocks also argue for an increase
in the rate of moderate and large earthquakes before a mainshock
[{\it Jaum\'e and Sykes}, 1999]. {\it Knopoff et al.} [1982] also suspected
a deviation from a linear Gutenberg-Richter distribution for foreshocks.
Our study of the ETAS model confirms that such a modification of the magnitude
distribution before a mainshock must be expected when averaging over many
foreshock sequences.

Intuitively, the modification of the magnitude distribution arises in our model
from the increase of the aftershock rate with the mainshock
magnitude. Any event has thus
a higher probability to occur just after a large event, because this
large event
induces an increase of the seismicity rate.
The novel properties that we demonstrate is that, before a mainshock,
the energy distribution is no more a pure power-law, but it is the sum
of the unconditional distribution
with exponent $\beta$ and an additional deviatoric power-law distribution with
a smaller exponent $\beta'=\beta- a$ as seen from expression (\ref{GRaalaw}).
In addition, we predict and verify numerically in figures \ref{fig6}
and \ref{fig7} that the amplitude of the deviatoric term increases as
a power-law of the time to the mainshock.
A similar behavior has been proposed as a precursory pattern termed
``pattern upward bend'' [{\it Keilis-Borok et al.}, 2001] or alternatively
providing ``pattern $\gamma$'' measured as the difference between the slope
of the Gutenberg-Richter for low and for large magnitudes. According
to our results,
pattern $\gamma$ should increase from $0$ to the value $a$.

	According to the ETAS model, the modification of the
magnitude distribution is independent
of the mainshock magnitude, as observed by
[{\it Kagan and Knopoff}, 1978; {\it Knopoff et al.} 1982;
{\it  Molchan and Dmitrieva}, 1990; {\it Molchan et al.}, 1999].
Therefore, all earthquakes, whatever their magnitude,
are preceded on average by an increase of the rate of large events.
Although the foreshock magnitude distribution is no more strictly speaking a
pure power-law but rather the sum of two power laws,
a single power-law  distribution with a decreasing
$b$-value as the time of the mainshock is approached
is a simple and robust way to quantify the increasing
importance of the tail of the distribution, especially for the short
foreshock sequences usually available. This rationalizes the suggestion
found in many works that a decrease in $b$-value is a (retrospective) signature
of an impending mainshock.
The novel insight provided by our analysis of the ETAS model is that a
better characterization of the magnitude distribution before mainshocks
may be provided by the sum of two power law distributions expressed by equation
(\ref{GRaalaw}) and tested in synthetic catalogs in Figures
\ref{fig6} and \ref{fig7}.
This rationalizes both the observed relatively small $b$-values reported for
foreshocks and the apparent decrease of $b$-value when the mainshock
approaches.
Similarly to the inverse Omori law, the modification of the
magnitude distribution
prior the mainshock is a statistical property which yields
an unambiguous signal only when stacking many foreshock sequences.
This may explain the variability of the patterns of $b$-value observed for
individual foreshock sequences.

A modification of the magnitude distribution before large earthquakes
is also expected from the critical point theory [{\it Sammis and
Sornette}, 2002].
The energy  distribution far from a critical point is characterized
by a power-law distribution with an exponential roll-off.
As the seismicity evolves towards the critical point, the truncation 
of the energy
distribution increases. At the critical point, the average energy
becomes infinite (in an infinite system)
and the energy distribution follows a pure power-law distribution.
This modification of the seismicity predicted by the critical point theory
is different from the one reported in this study, but the two models yield
an apparent decrease of $b$-value with the time from the mainshock. Therefore,
it is difficult to distinguish the two models in real seismicity data.
However, the difference between the two models is that
a modification of the energy distribution
should only be observed before major earthquakes according to
the critical point theory. Of course, one can not exclude that both
mechanisms occur and are mixed up in reality.

\subsection{Implications for earthquake prediction}

The inverse Omori law and the apparent decrease of $b$-value
have been derived in this study as statistical laws describing the average
fluctuations of seismicity conditioned on leading to a burst of
seismicity at the time of the mainshock.
This does not mean that there is not a genuine physical content in these
laws. We now demonstrate that they may actually embody an
important part of the physics of earthquakes and describe
the process of interactions between and triggering of earthquakes
by other earthquakes. For this purpose, we use
the modification of the magnitude frequency and the increase
of the seismicity rate as predicting tools of future individual mainshocks.
In the present work, we restrict our tests to the ETAS branching model
used as a playing ground for our ideas.

Using numerical simulations of the
ETAS model generated with $b=1$, $a=0.5 \beta$, $n=1$, $m_0=3$ and
$\theta=0.2$,
we find that large earthquakes occur more frequently following a
small locally estimated  $b$-value.
We have measured the $b$-value using a maximum likelihood method
for a sliding window of 100 events.
For instance, we find that 29\% of the large $M>6$ mainshocks
occur in a 11\% time period where $\beta$ is less than $95\%$ of the 
actual $b$-value (that is $b<0.95$). This leads to a significant
prediction gain of $g=2.7$, defined as the ratio of the successful
prediction (29\%)
over the duration of the alarms (11\%) [{\it Aki}, 1981].
A random prediction would lead $g=1$.

A much larger gain can be obtained using other precursory indicators related
to the inverse Omori law.
First, a large earthquake is likely to occur following another large
earthquake.
For the same simulation, fixing an alarm if the largest event within
the 100 preceding earthquakes is larger that $M=6$ yields a
probability gain $g=10$
for the prediction of a mainshock of magnitude equal to or larger than $M=6$.
Second, a large seismicity rate observed at a given ``present'' time
will lead on average to a large seismicity rate in the future, and thus
it increases the probability of having a large earthquake.
Measuring the seismicity rate over a sliding window
with flexible length imposed to contain exactly 100 events and fixing the alarm
threshold at $0.05$ events per day, we are able to predict 20\% of the
$M \geq 6$ events with just 0.16\% of the time period covered by the alarms.
This gives a prediction gain $g=129$.

Figure \ref{errordiagram} synthesizes and extends these results by
showing the so-called error diagram [{\it Molchan}, 1991; 1997]
for each of three functions measured in a sliding window of 100 events: (i) the
maximum magnitude $M_{max}$ of the 100 events
in that window, (ii) the apparent Gutenberg-Richter exponent $\beta$ measured
on these 100 events by the standard Hill maximum likelihood estimator and
(iii) the seismicity rate $r$ defined as the inverse of the duration
of the window containing 100 events.
For each function, an alarm is declared for the next event
when the function is either larger
(for $M_{max}$ and $r$) or smaller (for $\beta$) than a threshold.
Scanning all possible thresholds constructs the continuous curves shown in the
figure. The results on the prediction obtained by using these three
precursory functions
are considerably better than those obtained for a random prediction, shown
as a dashed line for reference. We have not tried
at all to optimize any facet of these prediction tests, which are offered
for the sole purpose of stressing the physical reality of the
precursory information contained in the foreshocks.

\subsection{Migration of foreshocks}

Among the proposed patterns of foreshocks, the migration of foreshocks
towards the mainshocks is much more difficult to observe
than either the inverse Omori law or the change in $b$-value. This is
due to the limited number of foreshocks and to the location errors.
Similarly to other foreshock patterns, a few case-histories have shown
seismicity migration before a mainshock. When reviewing 9 $M>7$ shallow
earthquakes in China, {\it Ma et al.} [1990] report a
migration of $M > 3-4$ earthquakes towards the mainshock
over a few years before the mainshock and at a distance of a few hundreds of
kilometers.
Less than 20 events are used for each case study.
While the case for the diffusion of aftershocks
is relatively strong [{\it Kagan and Knopoff},
1976, 1978; {\it von Seggern et al.}, 1981; {\it Tajima and Kanamori}, 1985]
but still controversial,
the migration of foreshocks towards the mainshock area, suggested
using a stacking method [e.g., {\it Kagan and Knopoff}, 1976, 1978;
{\it von Seggern et al.}, 1981; {\it Reasenberg}, 1985] is even less
clearly observed.

Using the ETAS model, {\it Helmstetter and Sornette} [2002b]
have shown that the cluster of aftershocks
diffuses on average from the mainshock according to the
diffusion law $R\sim t^H$, where $R$ is the typical size of the cluster
and $H$ is the so-called Hurst exponent which can be smaller
or larger than $1/2$.
In the present study, we have shown analytically and numerically that
this diffusion
of aftershocks must be reflected into a (reverse) migration of seismicity
towards the mainshock,
with the same diffusion exponent $H$ (defined in (\ref{H})).
We should however point out that
this predicted migration of foreshocks, as well as the diffusion of
aftershocks,
is significant only over a finite domain of the
parameter space over which the ETAS model is defined.
Specifically, a significant spatio-temporal coupling of the seismicity
leading to diffusion and migration is expected and observed in our
simulations only for
sufficiently large $\theta$'s and for short times $|t_c-t|<t^*$ from
the mainshock,
associated with a direct Omori exponent $p$ smaller than $1$.
This may explain why the diffusion of aftershocks and the migration
of foreshocks
is often difficult to observe in real data.

An additional difficulty in real
data arises from the background seismicity, which can induce a spurious
diffusion of aftershocks or migration of foreshocks (see Figure \ref{fig8}c).
As for the other foreshock patterns derived in this study,
the migration of foreshocks towards the mainshock and the spatial distribution
of foreshocks are independent of the mainshock magnitude. These results
disagree with the observations of [{\it Keilis-Borok and Malinovskaya}, 1964;
{\it Bowman et al.}, 1998] who suggest that
the area of accelerating seismicity prior a mainshock increases with
the mainshock
size. An increase of the foreshock zone with the mainshock size may
however be observed
in the ETAS model when using foreshocks
of type I (conditioned on being smaller than the mainshock)
and introducing a characteristic size of the aftershock zone $d$ in
(\ref{rhor}) increasing with the mainshock size.


\section{Conclusion}

We have shown that the ETAS (epidemic-type aftershock) branching model
of seismicity, based on the two best established empirical Omori and
Gutenberg-Richter laws, contains essentially all the phenomenology of
foreshocks. Using this model,
decades of empirical studies
on foreshocks are rationalized, including the inverse Omori law,
the $b$-value change and seismicity migration.
For each case, we have derived analytical solutions that relates
the foreshock distributions in the time, space and energy domain
to the properties of a simple earthquake triggering process embodied
by aftershocks. We find that all previously reported properties
of foreshocks arises from the Omori and Gutenberg-Richter law
when conditioning the spontaneous fluctuations of the rate of seismicity
to end with a burst of activity, which defines the time of the mainshock.
The foreshocks laws are seen as statistical laws which are clearly
observable when averaging over a large number
of sequences and should not be observed systematically when looking
at individual
foreshock sequences. Nevertheless, we have found that foreshocks contain
genuine important physical information of the triggering process
and may be used successfully to predict earthquakes with very significant
probability gains. Taking these results all together,
this suggests that the physics of aftershocks is sufficient
to explain the properties of foreshocks, and that there is no
essential physical
difference between foreshocks, aftershocks and mainshocks.

\vskip 0.5cm
\acknowledgments
We are very grateful to Y.Y. Kagan, G. Ouillon and
V. Pisarenko for useful suggestions and discussions,
and for R. Schoenberg for a constructive review of the manuscript. 
This work was partially supported by french INSU-Natural
Hazard grant (AH and JRG) and by the James S. Mc Donnell Foundation
21st century scientist award/studying complex system (DS).

\appendix
\section{Appendix A: Deviations from the average seismicity rate}

Using the definition of $\lambda (t)$ (\ref{maagmkr}), in the case
where the external $s(t)$
source term is a Dirac $\delta (t)$, we obtain the following expression for the
stochastic propagator
\be
\kappa(t)= \delta(t) +  \int \limits_{-\infty}^t d\tau~
\int \limits_{E_0}^{+\infty} dE~
\phi_{E}(t-\tau)  \sum_{i~|~t_i\leq t} \delta(E-E_{i})~
\delta(\tau-t_i)~,
\label{maaigmkr}
\ee

We now express the deviation of $\kappa(t)$ from its ensemble average $K(t)$.
This can be done by using (\ref{mgnnnmv}), which
means that the distribution density of earthquake energies is constructed
by recording all earthquakes and by counting the frequency of their energies.
Thus, $\delta(E-E_{\tau})$ can be seen as the sum of its average plus a
fluctuation part, namely, it
can be formally expressed as $\delta(E-E_{\tau})=P(E) + \delta P(E)$,
where $\delta P(E)$ denotes the fluctuation of $\delta(E-E_{\tau})$
around its ensemble average $P(E)$.
Similarly, $\kappa(t) = \sum_{t_i\leq t} \delta(t-t_i) =
K(t) + \delta \kappa(t)$, where $\delta \kappa(t)$ is the fluctuating
part of the seismic rate around its ensemble average $K(t)$.

We can thus express the sum of products of Dirac functions in
(\ref{maaigmkr}) as follows:
\be
\sum_{i~|~t_i\leq t} \delta(E-E_{i}) \delta(t-t_i) = P(E) K(t)
+\delta(P\kappa)(E,t)~.
\ee
As a first illustration, we can use the approximation that the fluctuations
of the product $\delta(E-E_{\tau}) \sum_{t_i\leq t} \delta(t-t_i)$
can be factorized to write
$$
\delta(E-E_{t}) \sum_{t_i\leq t} \delta(t-t_i) = \left(P(E) + \delta
P(E)\right) \left(K(t) + \delta \kappa(t)\right)
$$
\be
\approx P(E) K(t) + P(E) ~\delta \kappa(t) + K(t)~\delta P(E) ~.
\label{mghkr}
\ee
Using expression (\ref{maaigmkr}) for $\kappa(t)$ and expression
(\ref{thissrdter})
for $K(t)$, and putting (\ref{mghkr}) in (\ref{maaigmkr}), we then obtain
\be
\kappa(t) = K(t) +
\int \limits_{0}^t d\tau~ \int \limits_{E_0}^{+\infty} dE~
\phi_{E}(t-\tau) \delta(P\kappa)(E,\tau)~,
\label{maagmaasskr}
\ee
where
\be
\delta(P\kappa)(E,\tau) \equiv
\delta P(E) K(\tau) + P(E) \delta \kappa(t)~.
\label{ngjk}
\ee

By construction, the average of the double integral in the r.h.s. of
(\ref{maagmaasskr}) is zero. The double integral thus represents the
fluctuating part of the realization specific seismic response $\kappa(t)$ to
a triggering event.
Inserting (\ref{maagmaasskr}) in (\ref{mbnn}), we obtain
$$
\lambda(t) = N(t) +
$$
\be
\int \limits_{-\infty}^t d\tau ~s(\tau)
\int \limits_{0}^{t-\tau} du~\int \limits_{E_0}^{+\infty} dE~
\phi_{E}(t-\tau -u) \delta(P\kappa)(E,u)~.
\label{mghko}
\ee
Using $\int_{-\infty}^t d\tau  \int_{0}^{t-\tau} du =
\int_{0}^{+\infty} du  \int_{-\infty}^{t-u} d\tau$,
expression (\ref{mghko}) reads
$$
\lambda(t) = N(t) +
$$
\be
\int \limits_{E_0}^{+\infty} dE ~ \int \limits_{0}^{+\infty} du 
~\delta(P\kappa)(E,u)
        \int \limits_{-\infty}^{t-u} d\tau~ s(\tau) ~\phi_E(t-\tau -u)~.
        \label{nnwlgw}
\ee

For instance, let us consider the first contribution $\delta P(E) K(\tau)$ of
$\delta(P\kappa)(E,\tau)$ given by (\ref{ngjk}).
Denoting
\be
\epsilon \equiv \int \limits_{E_0}^{+\infty} dE ~\rho(E)~ \delta P(E)~,
\label{fmslla}
\ee
$\lambda(t)$ given by (\ref{nnwlgw}) is of the form (\ref{mgktsx})
with
\be
\eta(\tau) = \epsilon \int \limits_0^{+\infty} dx ~s(\tau -x)~\Psi(x)~,
\label{mgmmlr}
\ee
where $\Psi(x)$ is the bare Omori propagator defined in (\ref{mbjosl}).

The only property needed below is that the stochastic process
$\eta(\tau)$ be stationary. This is the case because the fluctuations
of $\delta P(E)$ and of the source $s(t)$ are stationary processes.
Similarly, the second contribution $P(E) \delta \kappa(\tau)$ of
$\delta(P\kappa)(E,\tau)$ given by (\ref{ngjk}) takes the form
(\ref{mgktsx}) if $\delta \kappa(\tau)$ is a noise
proportional to $K(t)$. At present, we cannot prove it but this
seems a natural assumption.
More generally, one could avoid the decomposition of
$\delta(P\kappa)(E,\tau)$ given by (\ref{ngjk}) and get the same
result as long as $\delta(P\kappa)(E,t)$ is equal to
a stationary noise multiplying $K(t)$.

\section{Appendix B: Conditioning weighted power law variables on the
realization of their sum}

Consider i.i.d. (identically independently distributed) random variables $x_i$
distributed according to a power law $p(x_i)$ with exponent $m \leq 2$.
Let us define the sum
\be
S_N=\sum_{i=1}^N K_i x_i~,
\ee
where the $K_i$'s are arbitrary positive weights.
Here, we derive that the expectation ${\rm E}[x_i|S_N]$ of $x_i$
conditioned on the existence of a large realization of $S_N$ is given
by (\ref{nbvkd}).

By definition, ${\rm E}[x_i|S_N] = N/D$ where
\be
N = \int dx_1~ ... \int dx_N~ x_i~p(x_1)...p(x_{N})~
\delta \left(S_N-\sum_{j=1}^N K_j x_j\right)~,
\label{mgnlw;a}
\ee
and $D$ is the same expression without the factor $x_i$. The Fourier
transform
of (\ref{mgnlw;a}) with respect to $S_N$ yields
\be
{\hat N}(k) = \left[\prod_{j \neq i} {\hat p}(k K_j)\right]~
{1 \over ik}~{d{\hat p}(k K_i)\over dK_i} =
{1 \over ik}~{d \over dK_i}\left[\prod_{j=1} {\hat p}(k K_j)\right]~.
\ee
We have used the identity
$\int dx_i~ x_i~p(x_i)~e^{ikK_i x_i} = {1 \over ik}~{d{\hat p}(k
K_i)\over dK_i}$
and ${\hat p}(k)$ is the Fourier transform of $p(x)$.
Note that $\prod_{j=1}^N {\hat p}(k K_j)$ is nothing but the Fourier
transform ${\hat P}_S(k)$ of the distribution $P_N(S_N)$. Using the
elementary identities of derivatives of Fourier transforms and by
taking the inverse Fourier transform, we thus get
\be
N = {d \over dK_i}\int \limits_{S_N}^{+\infty} dX ~P_N(X)~.
\ee
By definition, the denominator $D$ is identically equal to $P_N(S_N)$.
This yields the general result
\be
{\rm E}[x_i|S_N] = {1 \over P_N(S_N)}~{d \over
dK_i}\int \limits_{S_N}^{+\infty} dX ~P_N(X)~.
\label{mvawq}
\ee
In the special case where all $K_i$'s are equal, this gives the
``democratic''
result ${\rm E}[x_i|S_N]=S_N/N$.

For power law variables with distribution $p(x) \sim 1/x^{1+m}$ with
$m<2$, we can use the generalized central limit theorem to obtain that
$P_N(X)$ converges for large $N$ to a stable L\'evy law $L_m$ with
index equal to the exponent $m$ and scale factor $\sum_{j=1}^N K_j^m$
[{\it Gnedenko and Kolmogorov}, 1954; {\it Sornette}, 2000]:
\be
P_N(S_N) \to_{N\to \infty}  L_m\left({S_N \over \left(\sum_{j=1}^N
K_j^m\right)^{1/m}}\right)~.
\label{jgnlaa}
\ee
The only dependence of $P_N(S_N)$ in $K_i$ is found in the scale
factor. Putting the expression (\ref{jgnlaa}) into (\ref{mvawq}) yields the
announced result (\ref{nbvkd}). In particular, for $m=2$, this recovers the
standard result for Gaussian variables that ${\rm E}[x_i|S_N] \sim S_N K_i$,
because the stable L\'evy law of index $m=2$ is the Gaussian distribution.

\end{article}

\clearpage

\begin{figure}
\psfig{file=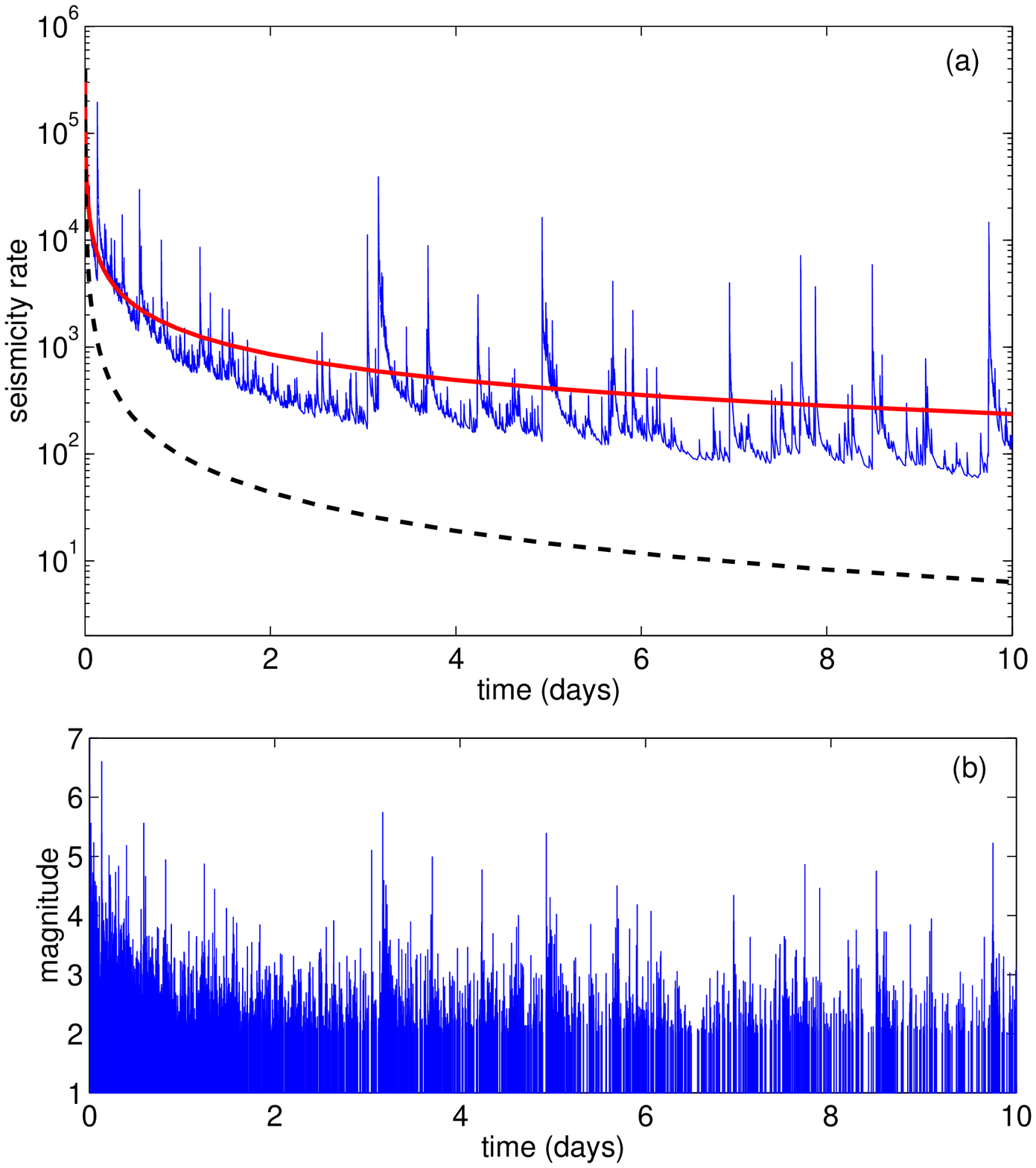,width=10cm}
\caption{\label{fig1} An example of a realization of the ETAS model, which
illustrates the differences between the observed seismicity rate $\kappa(t)$
(noisy solid line), the average renormalized (or dressed) propagator
$K(t)$ (solid line),
and the local propagator $\phi_E(t)$ (dashed line).
The magnitude of each earthquake are shown in panel (b).
This aftershock sequence has been generated using the ETAS model with
parameters $n=1$, $a=0.8\beta$, $\theta=0.2$, $m_0=2$ and $c=0.001$
day, starting from a
mainshock of magnitude $M=7$ at time $t=0$. The global aftershock rate
$\kappa(t)$ is significantly higher than the direct (or first
generation) aftershock rate,
described by the local propagator $\phi_E(t)$.
The global aftershock rate $\kappa(t)$
decreases on average according to the dressed propagator $K(t) \sim
1/t^{1-\theta}$,
which is significantly slower than
the local propagator $\phi(t) \sim 1/t^{1+\theta}$. The best fit to
the observed
seismicity rate $\kappa(t)$ is indistinguishable from the average dressed
propagator $K(t)$.
Large fluctuations of the seismicity rate corresponds to the
occurrence of large
aftershocks, which trigger their own aftershock sequence.
Third-generation aftershocks
can be easily observed.}
\end{figure}

\clearpage
\begin{figure}
\psfig{file=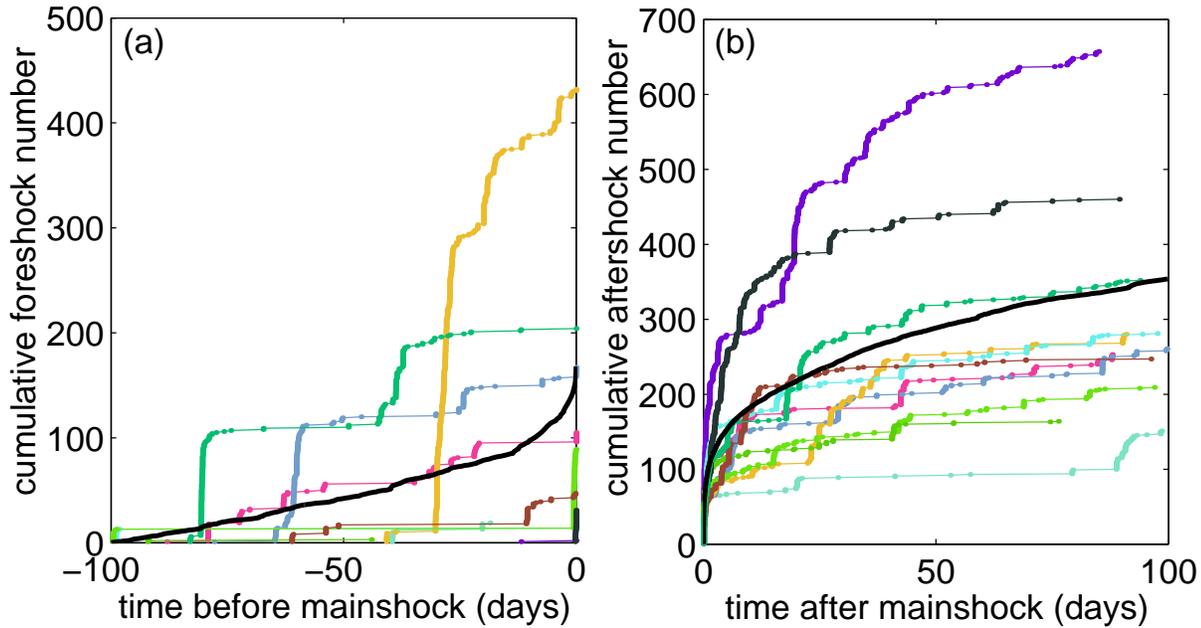,width=16cm}
\caption{\label{fig2} Typical foreshock (a) and aftershock (b) sequences
generated by the ETAS model, for mainshocks of magnitude $M=5.5$ occurring
at time $t=0$.
We show 11 individual sequences in each panel.
The solid black line represents
the mean seismicity rate before and after a mainshock of magnitude $M=5.5$,
estimated by averaging over 250 sequences.
The synthetic catalogs have been generated using the parameters $n=1$,
$\theta=0.2$, and $a=0.5 \beta$, with a minimum magnitude threshold $m_0=2$.
In contrast with the direct Omori law, which is clearly observed
after any large mainshock, there are large fluctuations from one foreshock
sequence to another one, and the inverse Omori law (with accelerating
seismicity) is only observed
when averaging over a large number of foreshock sequences.
}
\end{figure}

\begin{figure}
\psfig{file=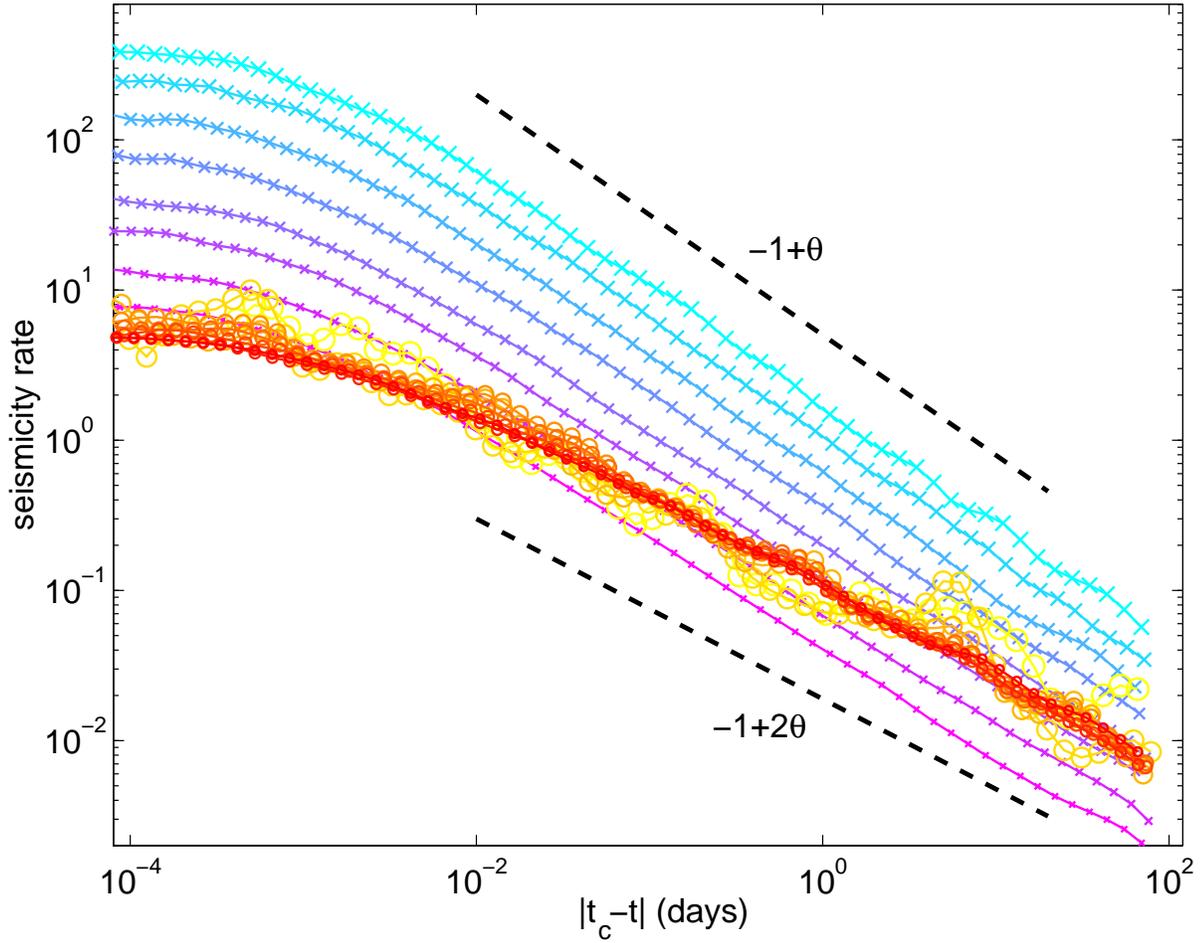,width=16cm}
\caption{\label{fig3} Direct and inverse Omori law for a numerical simulation
with $a=0.5 \beta$ and $\theta=0.2$ showing the two exponents $p=1-\theta$
for aftershocks and $p'=1-2\theta$ for foreshocks of type II.
The rate of aftershocks (crosses) and foreshocks (circles) per mainshock,
averaged over a large number of sequences, is shown
as a function of the time  $|t_c-t|$ to the mainshock, for different values
of the mainshock magnitude between 1.5 and 5, with a step of 0.5.
The symbol size increases with the mainshock magnitude. The truncation of
the seismicity rate for small times $|t_c-t|\simeq 0.001$ day is due to the
characteristic time $c=0.001$ day in the bare Omori propagator $\Psi(t)$,
and is the same for foreshocks and aftershocks.
The number of aftershocks increases with the mainshock energy as $N\simeq E^a$,
whereas the number of foreshocks of type II in independent of the
mainshock energy.
}
\end{figure}

\clearpage

\begin{figure}
\psfig{file=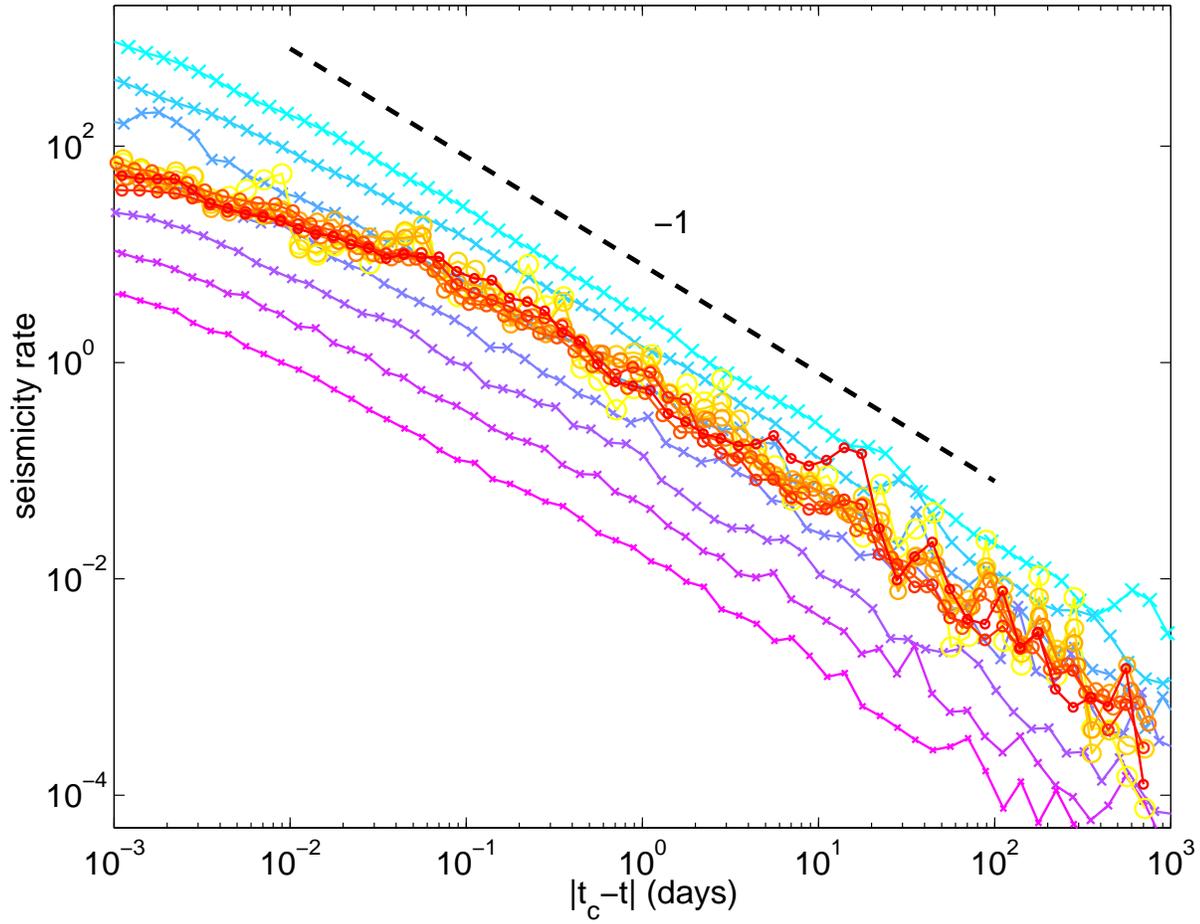,width=16cm}
\caption{\label{fig4} Same as Figure \ref{fig3} for $a=0.8 \beta$,
showing the larger relative ratio of foreshocks to aftershocks
compared to the case $a=0.5 \beta$.
}
\end{figure}

\begin{figure}
\psfig{file=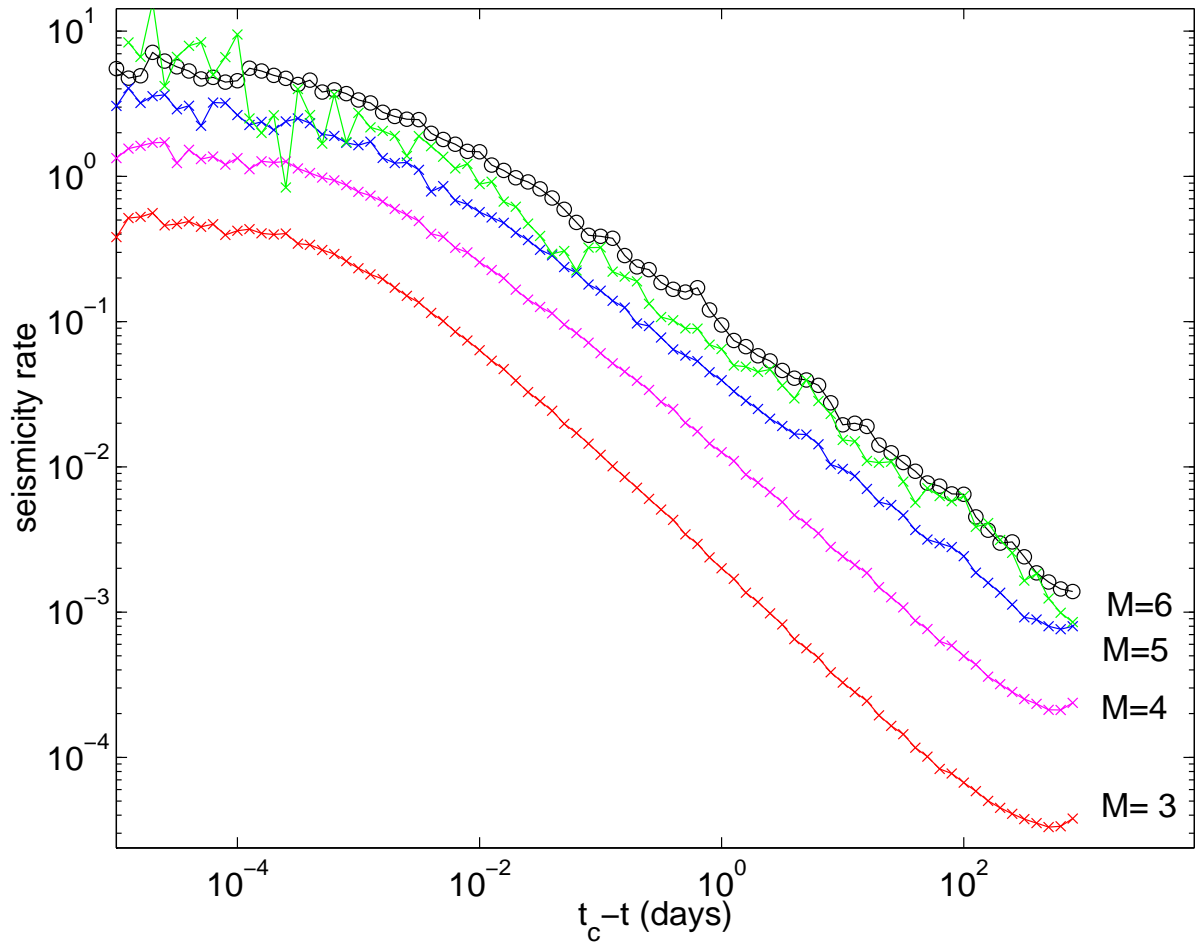,width=16cm}
\caption{\label{typeI} Foreshock seismicity rate per mainshock
for foreshocks of type II (circles)
and foreshocks of type I (crosses), for a numerical simulation with $n=1$,
$c=0.001$ day, $\theta=0.2$, $a=0.5\beta$ and $m_0=2$.
For foreshocks of type I, we have considered mainshock magnitudes $M$ ranging
from $3$ to $6$. We have rejected from the analysis of foreshocks of type I all
mainshocks which have been preceded  by a larger event in a time interval
extending up to $t=1000$ days preceding the mainshock.
The rate of foreshocks of type II is independent on the mainshock
magnitude $M$,
while the rate of foreshocks of type I increases with $M$. For large
mainshock magnitudes, the rate of foreshocks of type I is very close to
that of foreshocks of type II.
The conditioning that foreshocks of type I must be smaller than their
mainshock induces an apparent increase of the Omori exponent $p'$ as
the mainshock magnitude decreases. It induces also an upward bending of
the seismicity rate at times $t\approx 1000$ days, especially for the small
magnitudes.
}
\end{figure}
\clearpage

\clearpage
\begin{figure}
\psfig{file=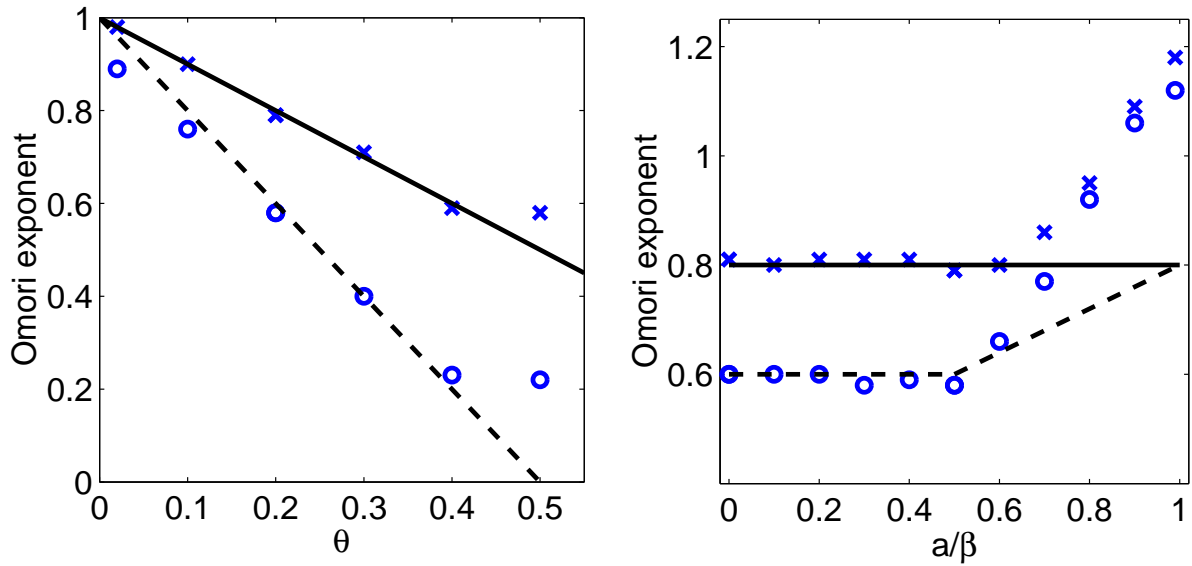,width=16cm}
\caption{\label{fig5} Exponents $p'$ and $p$ of the inverse and
direct Omori laws
obtained from numerical simulations of the ETAS model.
The estimated values of $p'$ (circles) for foreshocks and $p$ (crosses)
for aftershocks are shown as a function of $\theta$ in the case
$\alpha=0.5$ (a),
and as a function of $a/\beta$ in the case $\theta=0.2$ (b).
For $a/\beta$ not too large, the values of $p'$ for foreshocks are in
good agreement
with the predictions $p'=1-2\theta$ for $a/\beta<0.5$ (\ref{mgmlks}) and
$p'=1-\beta ~\theta/a$ for $a/\beta>0.5$ (\ref{pmse}).
The theoretical values of $p'$ are represented with dashed lines in each plot,
and the theoretical prediction for $p$ is shown as solid lines.
For $a/\beta$ not too large, the measured exponent for aftershocks is in good
agreement with the prediction $p=1-\theta$ (\ref{soklfm}).
For $a/\beta>0.5$, both $p$ and $p'$-values are larger than the predictions
  (\ref{soklfm}) and (\ref{pmse}).
For $a/\beta$ close to 1, both $p$ and $p'$ are found close to the exponent
$1+\theta=1.2$ of the bare propagator $\psi(t)$. See text for an explanation.}
\end{figure}

\clearpage

\begin{figure}
\psfig{file=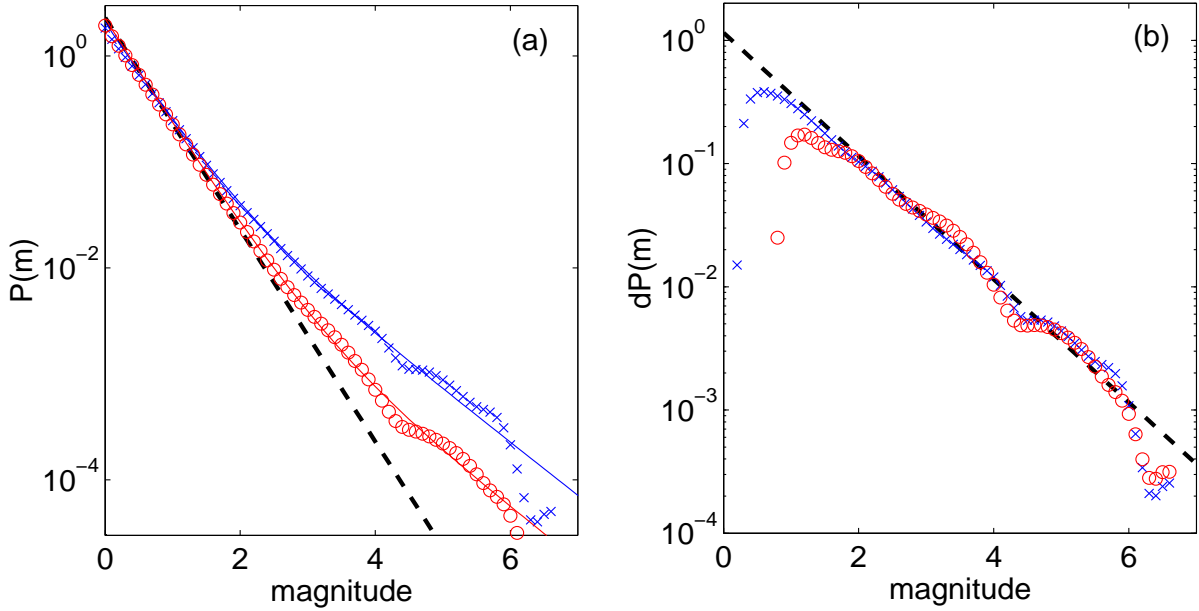,width=16cm} 
\caption{\label{fig6} Magnitude distribution of foreshocks for two
time periods: $t_c-t<0.1$ days (crosses) and $1<t_c-t<10$ days (circles), 
for a numerical simulation
of the ETAS model with parameters $\theta=0.2$, $\beta=2/3$,
$c=10^{-3}$ day, $m_0=2$ and $a=\beta/2=1/3$.
The magnitude distribution $P(m)$ shown on the first plot (a)
has been build by stacking many foreshock sequences of magnitudes
$M>4.5$ mainshocks.
The observed magnitude distribution is in very good agreement with
the prediction (\ref{GRaalaw}), shown as a solid line for each time period,
that the magnitude distribution is the sum of the unconditional
Gutenberg-Richter law with an exponent $b=1.5\beta=1$, shown as a
dashed black line, and a deviatoric Gutenberg-Richter law $dP(m)$
with an exponent $b'=b-\alpha=0.5$ with $\alpha=1.5a=0.5$.
The amplitude of the perturbation increases if $t_c-t$ decreases as expected
from (\ref{GRaalaw}).
The observed deviatoric magnitude distribution $dP(m)$ is shown on plot (b)
for the same time periods, and is in very good agreement with the
prediction shown as a dashed black line.
We must stress that the energy distribution is {\it no more} a pure
power law close to the mainshock, but the sum of two power laws.
The panel on the right exhibits the second power law which is created
by the conditioning mechanism underlying the appearance of
foreshocks. See text.}
\end{figure}

\clearpage

\begin{figure}
\psfig{file=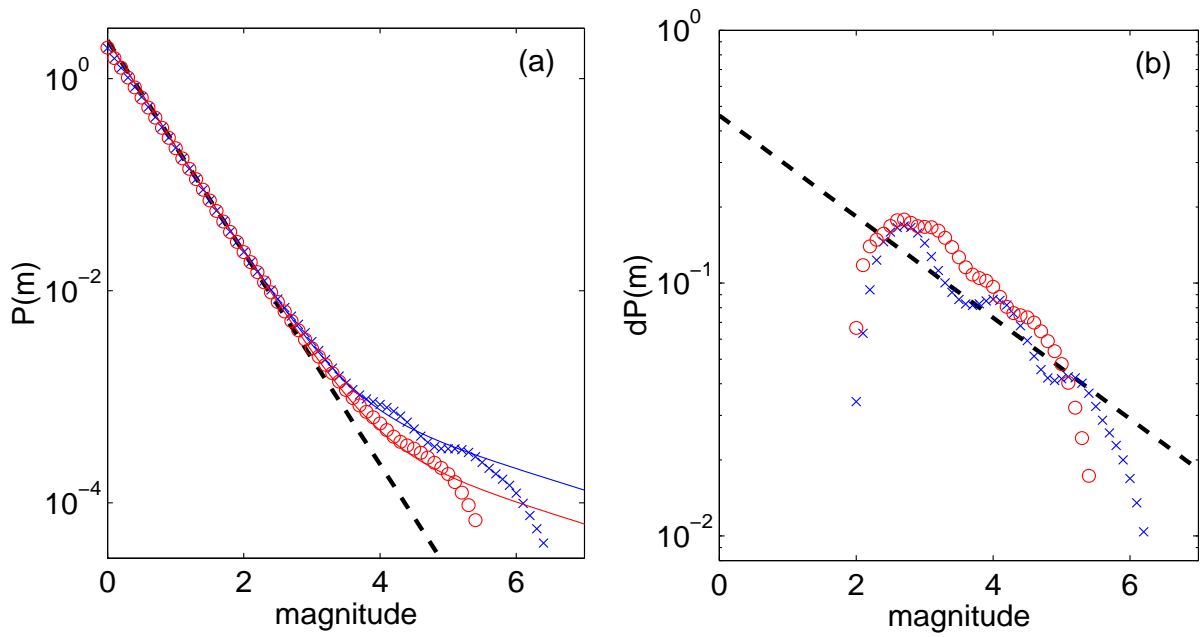,width=16cm} 
\caption{\label{fig7} Same as Figure \ref{fig6} but for $a=0.8 \beta$.
In this case, the deviatoric Gutenberg-Richter contribution is observed only
for the largest magnitudes, for which the statistics is the poorest, hence the
relatively large fluctuations around the exact theoretical predictions.
}
\end{figure}

\begin{figure}
\psfig{file=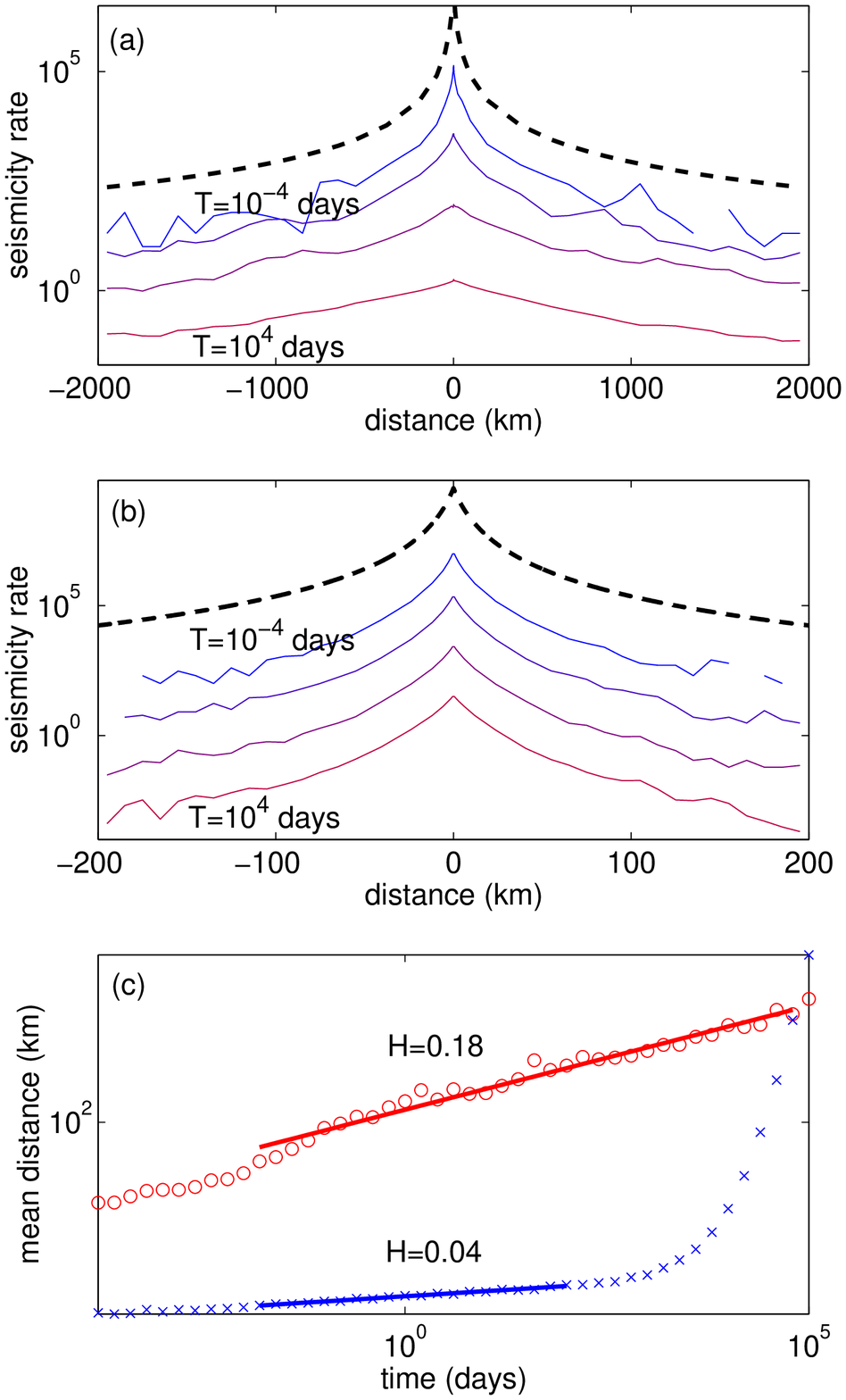,width=14cm}
\caption{\label{fig8} Migration of foreshocks, for superposed foreshock
sequences generated with the ETAS model for two choices of parameters,
(a) $n=1$, $\theta=0.2$, $a=0.5\beta$, $\mu=1$, $d=10$ km, $c=0.001$ day, 
$m_0=2$ and   (b) $n=1$, $\theta=0.02$, $a=0.5\beta$, $\mu=3$, $d=1$ km, 
$c=0.001$ day, $m_0=2$.
The distribution of foreshock-mainshock distances is shown on panel (a) and (b)
for the two simulations, for different time periods ranging between
   $10^{-4}$ to $10^4$ days. The distribution of mainshock-aftershock
   distances given by (\ref{rhor}) describing direct lineage
    is shown as a dashed line for reference.
On panel (a), we see clearly a migration of the seismicity towards
the mainshock,
as expected by the significant diffusion exponent $H=0.2$ predicted by
(\ref{H}). In contrast, the distribution of the foreshock-mainshock distances
shown in panel (b) is independent of the time from the mainshock,
as expected by the much smaller exponent diffusion $H=0.01$
predicted by (\ref{H}).
The characteristic size of the foreshock cluster is shown as a
function of the time
to the mainshock on panel (c) for the two numerical simulations.
Circles correspond to the simulation shown in panel (a) and crosses correspond
to the simulation shown in panel (b). The solid line is a fit of the
characteristic size of the foreshock cluster by $R \sim t^H$. For the
simulation
generated with $\theta=0.2$ and $\mu=1$ (circles), we obtain $H=0.18
\pm 0.02$ in very
good agreement with the prediction $H=\theta/\mu=0.2$ (\ref{H}).
The simulation generated with $\theta=0.02$ and $\mu=3$ (crosses) has a
much smaller exponent $H=0.04 \pm 0.02$, in good agreement with the
expected value
$H=\theta/2=0.01$ (\ref{H}). A faster apparent migration is observed
at large times for this simulation,
   due to the transition from the uniform background distribution for large
   times preceding the mainshock to the clustered seismicity prior to
the mainshock.
}
\end{figure}

\clearpage

\begin{figure}
\psfig{file=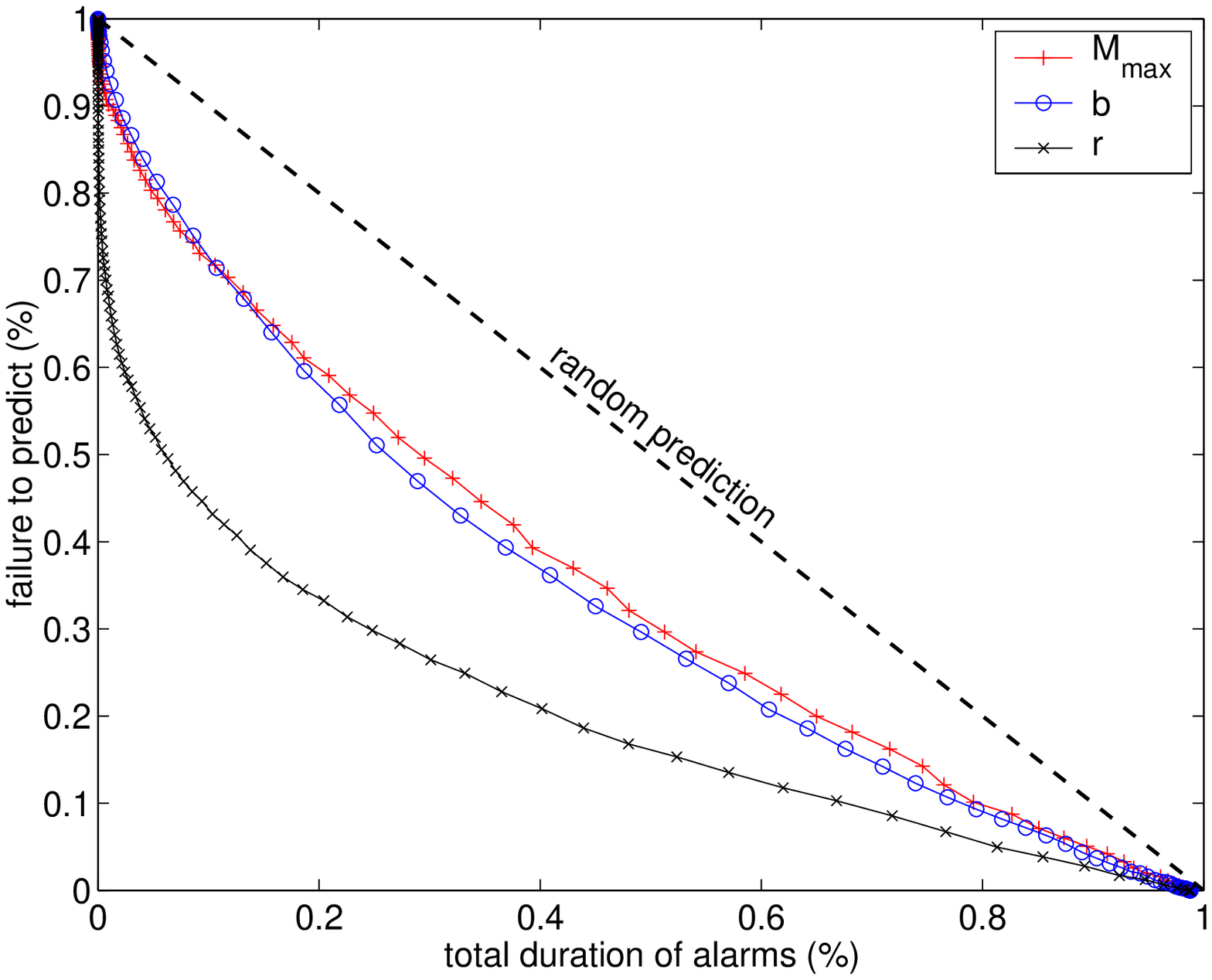,width=14cm}
\caption{\label{errordiagram}
Results of prediction tests for synthetic catalogs generated with
the parameters $a=0.5\beta$, $n=1$, $\beta=2/3$, $\theta=0.2$, $c=0.001$ day 
and a constant source $\mu=0.001$ shocks per day. 
The minimum magnitude is $m_0=3$ and the
target events are $M\geq 6$ mainshocks. We have generated 500
synthetics catalogs of 10000 events each,
leading to a total of 4735 $M\geq 6 $ mainshocks.
We use three functions measured in a sliding window of 100 events: (i) the
maximum magnitude $M_{max}$ of the 100 events
in that window, (ii) the apparent Gutenberg-Richter exponent $\beta$ measured
on these 100 events by the standard Hill maximum likelihood estimator and
(iii) the seismicity rate $r$ defined as the inverse of the duration
of the window.
For each function, we declare an alarm when the function is either larger
(for $M_{max}$ and $r$) or smaller (for $\beta$) than a threshold.
Once triggered, each alarm remains active as long as the function remains
larger (for $M_{max}$ and $r$) or smaller (for $\beta$) than the threshold.
Scanning all possible thresholds constructs the continuous curves shown in the
error diagram.
The quality of the predictions is measured by plotting
the ratio of failures to predict
as a function of the total durations of the alarms
normalized by the duration of the catalog. The results for these
three functions
are considerably better than those obtained for a random prediction, shown
as a dashed line for reference. The best results are obtained using the
seismicity rate. Predictions based on the Gutenberg-Richter
$\beta$ and on the maximum magnitude observed within the running window provide
similar results.
}
\end{figure}

\begin{references}




\reference Abercrombie, R. E., and J. Mori, 
	Occurrence patterns of foreshocks to
	large earthquakes in the western United States,
	{\it Nature, 381}, 303-307, 1996.

\reference Agnew, D. C. and L. M. Jones, Prediction probabilities
from foreshocks,
	{\it J. Geophys. Res., 96}, 11,959-11971, 1991.

\reference Aki, K., A probabilistic synthesis of precursory phenomena, in
	{\it Earthquake Prediction}, edited by D. W. Simpson and P. 
G. Richards,
	 AGU Maurice Ewing series: 4, Washington, D.C., pp 556-574, 1981.

\reference Berg, E., Relation between earthquake foreshocks, stress
and mainshocks,
	{\it Nature, 219}, 1141-1143, 1968.

\reference Bowman, D. D., G. Ouillon, C. G. Sammis, A. Sornette and 
D. Sornette,
	An observational test of the critical earthquake concept,
	{\it J. Geophys. Res.}, {\it 103}, 24359-24372, 1998).

\reference Bufe, C. G., Frequency magnitude variations during the 1970
Danville earthquake swarm,
	{\it Earthquake Notes, 41}, 3-7, 1970.

\reference Cagnetti, V. and V. Pasquale, The earthquake sequence in
Frieli, Italy, 1976,
             {\it  Bull. Seism. Soc. Am.}, {\it 69}, 1797-1818, 1979

\reference Console, R., M. Murru and B. Alessandrini, Foreshocks
statistics and their possible
relationship to earthquake prediction in the Italian region,
{\it  Bull. Seism. Soc. Am.}, {\it 83}, 1248-1263, 1983.

\reference Daley, D. J. and D. Vere-Jones, {\it An Introduction to
	the theory of point processes},  Springer, 1988.

\reference Davis, S. D. and C. Frohlich, Single-link cluster analysis of
              earthquake aftershocks: decay laws and regional variations,
               {\it J. Geophys. Res.}, {\it 96}, 6335-6350, 1991.

\reference Dodge, D. D., G. C. Beroza and W.L. Ellsworth,
	Detailed observations of California foreshock sequence:
	Implications for the earthquake initiation process,
         {\it J. Geophys. Res., 101}, 22,371-22,392, 1996.

\reference Enescu, B. and K. Ito,
	Some premonitory phenomena of the 1995 Hyogo-Ken Nanbu (Kobe)
	earthquake: seismicity, $b$-value and fractal dimension,
         {\it Tectonophysics}, {\it 338}, 297-314, 2001.


\reference Fedotov S. A., A. A. Gusev and S. A. Boldyrev, Progress in
    earthquake prediction in Kamchatka,
         {\it Tectonophysics}, {\it 14}, 279-286, 1972.



\reference Felzer, K. R., T. W. Becker, R. E. Abercrombie, G. Ekstr\"om
    and J. R. Rice, Triggering of the 1999 $M_W$ $7.1$ Hector Mine
    earthquake by aftershocks of the 1992 $M_W$ $7.3$ Landers
    earthquake, {\it J. Geophys. Res.}, 107, 2190,
    doi:10.1029/2001JB000911, 2002. 

\reference Gabrielov, A., I.  Zaliapin, W. I. Newman and V. I. Keilis-Borok,
	Colliding cascades model for earthquake prediction,
	{\it Geophys. J. Int.},{\it 143}, 427-437, 2000a.

\reference Gabrielov, A., V. Keilis-Borok, I. Zaliapin and W. I. Newman,
	Critical transitions in colliding cascades,
	{\it Phys. Rev. E}, {\it 62}, 237-249, 2000b.

\reference Gnedenko, B. V. and A. N. Kolmogorov,  {\it Limit
	Distributions for Sum of Independent Random Variables}, Addison
	Wesley,~Reading MA, 1954.

\reference Gross, S. and C. Kisslinger, Tests of models of aftershocks rate
         decay, {\it  Bull. Seism. Soc. Am.}, {\it 84}, 1571-1579, 1994.

\reference Guo, Z. and Y. Ogata, Statistical relations between the
      parameters of aftershocks in time, space and magnitude,
      {\it J. Geophys. Res.}, {\it 102}, 2857-2873, 1997.

\reference Hainzl, S., G. Zoller and J. Kurths, Similar power laws
for foreshock
	and aftershock sequences in a spring-block model for earthquakes,
	{\it J. Geophys. Res., 104}, 7243-7253, 1999.

\reference Harris, T.E., {\it The theory of branching processes},
           Springer, Berlin, 1963.

\reference Helmstetter, A., Is earthquake triggering driven by 
small earthquakes?,
	{\it Phys. Res. Lett.},  91, 058501, 2003

\reference Helmstetter, A. and D. Sornette,
              Sub-critical and super-critical regimes in epidemic models of
              Earthquake Aftershocks, {\it J. Geophys. Res.}, 107,
              2237, doi:10.1029/2001JB001580, 2002a.

\reference Helmstetter, A. and D. Sornette, Diffusion of earthquake
aftershock epicenters and Omori law: exact mapping to generalized
continuous-time random walk models,
	{\it Phys. Rev. E}, 6606, 1104, doi:10.1103/PhysRevE.66.061104, 2002b. 

\reference Huang, Y., H. Saleur, C. Sammis and D. Sornette, Precursors,
	aftershocks, criticality and self-organized criticality,
	{\it Europhys. Lett., 41}, 43-48, 1998.

\reference Ikegami, R, {\it Bull. Earth. Res. Inst., 45}, 328-345, 1967.

\reference Imoto, M., Changes in the magnitude frequency $b$-value
	prior to large (M$\geq6.0$) earthquakes in Japan,
	{\it Tectonophysics}, {\it 193}, 311-325, 1991.

\reference Jacod, J. and  A. N. Shiryaev,
	{\it Limit Theorems for Stochastic Processes}, Springer, Berlin, 1987.

\reference Jaum\'e, S. C. and L. R. Sykes, Evolving towards a critical point:
	A review of accelerating seismic moment/energy release prior
to large and
	great earthquakes, {\it Pure Appl. Geophys., 155}, 279-306, 1999.

\reference Jones, L. M., Foreshocks (1966-1980) in the San Andreas system,
	California, {\it Bull. Seis. Soc. Am., 74}, 1361-1380, 1984.

\reference Jones, L. M.,  and P. Molnar, Some characteristics of foreshocks and
	their possible relationship to earthquake prediction and premonitory
	slip on fault, {\it J. Geophys. Res.}, {\it 84}, 3596-3608, 1979.

\reference Jones, L. M., R. Console, F. Di Luccio and M. Murru, Are
	foreshocks mainshocks whose aftershocks happen to be big?
	preprint 1999 available at\\
	http://pasadena.wr.usgs.gov/office/jones/italy-bssa.html

\reference Kagan, Y. Y., Aftershock zone scaling, {\it Bull. Seism.
	Soc. Am.}, {\it 92}, 641-655, 2002.

\reference Kagan, Y. Y. and L. Knopoff, Statistical search for
	non-random features of the seismicity of strong earthquakes,
	{\it Phys. Earth Planet. Int.}, {\it 12}, 291-318, 1976.

\reference Kagan, Y. Y. and L. Knopoff, Statistical study of the
	occurrence of shallow earthquakes, {\it Geophys. J. R. Astr. Soc.},
	{\it 55}, 67-86, 1978.

\reference Kagan, Y. Y. and L. Knopoff,
              Stochastic synthesis of earthquake catalogs,
              {\it  J. Geophys. Res.}, {\it 86}, 2853-2862, 1981.

\reference Kagan, Y. Y. and L. Knopoff,
                 Statistical short-term earthquake prediction,
                 {\it Science}, {\it 236}, 1563-1467, 1987.

\reference Keilis-Borok, V. I. and L. N. Malinovskaya,
	One regularity in the occurrence of strong earthquakes,
	 {\it  J. Geophys. Res.}, {\it 69}, 3019-3024, 1964.

\reference Keilis-Borok, V. and  V. G. Kossobokov,
	Premonitory activation of earthquake flow - Algorithm M8,
	{\it Phys. Earth Planet. Int.}, {\it 61}, 73-83, 1990.

\reference Keilis-Borok, V., A. Ismail-Zadeh, V. Kossobokov and P. Shebalin,
           Non-linear dynamics of the lithosphere and intermediate-term
	earthquake prediction, {\it Tectonophysics}, {\it 338}, 247-260, 2001.

\reference Kisslinger, C., The stretched exponential function as an
	alternative model for	aftershock decay rate,
	 {\it  J. Geophys. Res.}, {\it  98}, 1913-1921, 1993.

\reference Kisslinger, C. and L. M. Jones, Properties of aftershocks
	sequences in Southern California, {\it  J. Geophys. Res.},
	{\it   96}, 11947-11958, 1991.

\reference Knopoff, L., Y. Y. Kagan and R. Knopoff, $b$-values for
              fore- and aftershocks in real and simulated earthquakes sequences,
              {\it   Bull. Seism. Soc. Am.},  {\it  72}, 1663-1676, 1982.

\reference Li, Q. L., J. B. Chen, L. Yu, and B. L. Hao, Time and
	space scanning of
	the $b$-value: A method for monitoring the development of catastrophic
	earthquakes, {\it Acta Geophys. Sinica, 21}, 101-125, 1978.

\reference Lindh, A. G., and M. R. Lim, A clarification, correction
	and updating of
	Parkfield California, earthquakes prediction scenarios and response
	plans, {\it U.S. Geol. Surv. Open File Rep.}, 95-695, 1995.


\reference  Ma, H., Variation of the $b$-values before several large
	earthquakes that occurred in North China,
  	{\it Acta Geophysica Sinica., 21}, 126-141, 1978, (in chinese).

\reference  Ma, Z., Fu Z., Zhang Y., Wang C., Zhang G. and D. Liu,
	{\it Earthquake prediction, nine major cas in China},
  	 Seismological Press Beijing, Springer Verlag, 332 pp, 1990.

\reference Maeda, K., Time distribution of immediate foreshocks obtained by a
	stacking method, {\it Pure Appl. Geophys., 155}, 381-384, 1999.

\reference Michael, A. J. and L. M. Jones, Seismicity alert probabilities at
	Parkfield, California, revisited, {\it Bull. Seis. Soc. Am., 88},
	117-130, 1998.

\reference Mogi, K. Some discussions on aftershocks, foreshocks and
earthquake swarms,
    {\it Bull. Res. Inst., Tokyo, Univ.}, {\it 41}, 595-614, 1963.

\reference Mogi, K. Earthquakes and fractures,  {\it Tectonophysics},
{\it 5}, 35-55, 1967.

\reference  Molchan, G. M., Structure of optimal strategies in earthquake
prediction, {\it Tectonophysics}, {\it 193}, 267-276, 1991.

\reference  Molchan, G. M., Earthquake prediction as a decision-making problem,
{\it Pure Appl. Geophys.}, {\it 149}, 233-247, 1997.

\reference Molchan, G. M. and O. Dmitrieva, Dynamics of the magnitude
frequency relation for foreshocks
          {\it Phys. Earth. Plan. Inter.}, {\it 61}, 99-112, 1990.

\reference Molchan, G. M., T.L. Konrod and A.K. Nekrasova, Immediate
	foreshocks: time variation of the $b$-value,
          {\it Phys. Earth. Plan. Inter.}, {\it 111}, 229-240, 1999.

\reference Morse, P. M. and H. Feshbach, {\it Methods in Theoretical
	Physics}, McGraw-Hill, New-York, 1953.

\reference Narteau, C., P. Shebalin, M., Holschneider, J. L. Le Mouel and
C. All\`egre, Direct simulations of the stress redistribution in the scaling
organization of fracture tectonics (SOFT) model,
{\it Geophys. J. Int.}, {\it 141}, 115-135, 2000.

\reference Ogata, Y., Statistical models for earthquake occurrence
              and residual analysis for point processes, {\it J. Am.
	stat. Assoc.}, {\it  83}, 9-27, 1988.

\reference Ogata, Y., T. Utsu and K. Katsura, Statistical features of
	foreshocks in comparison with others earthquakes clusters,
	 {\it Geophys. J. Int., 121}, 233-254, 1995.

\reference Ogata, Y., T. Utsu and K. Katsura, Statistical discrimination of
	foreshocks from other earthquakes clusters, {\it Geophys. J.
	Int., 127}, 17-30, 1996.

   \reference  Omori, F., On the aftershocks of earthquakes,
           {\it J. Coll. Sci. Imp. Univ. Tokyo}, {\it 7}, 111-120, 1894.

\reference  Omori, F., On  Foreshocks of earthquakes,
           {\it Pub. Imp. Eartq. Inv. Com.}, {\it 2}, 89-100, 1908.

\reference Page, R.A., Comments on "earthquake frequency and
	prediction" by Liu Z.R.,
   	{\it Bull. Seis. Soc. Am.}, {\it 74}, 1491-1496, 1986.


\reference Papazachos, B. C., The time distribution of reservoir-associated
	foreshocks and its importance to the prediction of the principal shock,
	{\it Bull. Seis. Soc. Am., 63}, 1973-1978, 1973.

\reference Papazachos, B. C., On certain aftershock and foreshock
	parameters in the area of Greece, {\it Ann., Geofis., 28},
	497-515, 1975a.

\reference Papazachos, B., Foreshocks and earthquakes prediction,
	{\it Tectonophysics, 28}, 213-216, 1975b.

\reference Papazachos, B., M. Delibasis, N. Liapis, G. Moumoulis and G.
	Purcaru, Aftershock sequences of some large earthquakes in
	the region of Greece, {\it Ann. Geofis., 20}, 1-93, 1967.

\reference Pelletier, J. D., Spring-block models of seismicity/
	Review and analysis of a structurally heterogeneous model
	coupled to a viscous asthenosphere,
	in {\it Geocomplexity and the physics of earthquakes},
	edited by J.B. Rundle, D. L. Turcotte, and W. Klein, AGU,
	geophysical monograph 120, Washington D.C., 27-42, 2000.

\reference Reasenberg P., Second order moment of central California seismicity,
	1969-1982, {\it J. Geophys. Res., 90}, 5479-5495, 1985.

\reference Reasenberg, P. A., Foreshocks occurrence before large
	earthquakes, {\it J. Geophys. Res.}, {\it 104}, 4755-4768, 1999.

\reference Reasenberg, P. A. and L. M. Jones, earthquake hazard after a
              mainshock in California, {\it  Science}, {\it 243},
1173-1176, 1989.

\reference Richter, C. F., {\it Elementary seismology}, 758 pp.,
	 W.H. Freeman and Co, San Francisco, 1968.

\reference Rotwain, I., V. Keilis-Borok and L. Botvina,
            Premonitory transformation of steel fracturing and seismicity,
            {\it Phys. Earth Planet. Int.}, {\it 101}, 61-71, 1997.

\reference Sammis, S. G. and D. Sornette,
	Positive feedback, memory and the predictability of earthquakes,
	{\it Proceedings of the National Academy of Sciences},
	{\it 99}, 2501-2508, 2002.

\reference Scholz, C. H., Microfractures, aftershocks, and seismicity,
	{\it Seism. Soc. of Am. Bull.}, {\it 58}, 1117-1130, 1968.

\reference von Seggern, D., S. S. Alexander and C-B Baag,
Seismicity parameters preceding moderate to major earthquakes
{\it J. Geophys. Res., 86}, 9325-9351, 1981.

\reference Shaw, B. E., Generalized Omori law for aftershocks and
	foreshocks from a simple  dynamics,
	{\it Geophys. Res. Letts.,10}, 907-910, 1993.

\reference Smith, W.D., The $b$-value as an earthquake precursor,
                 {\it Nature, 289}, 131-139,1981.

\reference Smith, W.D., Evidence for precursory changes in the
frequency-magnitude $b$-value,
{\it Geophys., J. Roy. Astr. Soc., 86}, 815-838, 1986.

\reference Sornette, A. and D. Sornette, Renormalization of
              earthquake aftershocks,
              {\it Geophys. Res. Lett.}, {\it 26}, 1981-1984, 1999.

\reference Sornette, D., {\it Critical Phenomena in Natural Sciences,
              Chaos, Fractals, Self-organization and Disorder: Concepts
and Tools}, Springer Series in Synergetics, Heidelberg, 2000.

\reference Sornette D. and A. Helmstetter, 
On the occurrence of finite-time-singularities in
epidemic models of rupture, earthquakes and starquakes, 
{\it Phys. Rev. Lett., 89}, 158501, 2002 
(http://arXiv.org/abs/cond-mat/0112043).


\reference Sornette, D. and C.G. Sammis, Complex critical
exponents from renormalization group theory of earthquakes :
Implications for earthquake
predictions, {\it J. Phys. I France}, {\it 5}, 607-619, 1995.

\reference Sornette, D., C. Vanneste and L. Knopoff,
	Statistical model of earthquake foreshocks, {\it Phys. Rev. A},
	{\it 45}, 8351-8357, 1992.

\reference Stephens, C. D., J. C. Lahr, K. A. Fogleman and R.B. Horner,
	The St-Helias, Alaska, earthquake of february 28, 1979:
	regional recording of aftershocks and short-term, 
pre-earthquake seismicity, 
   	{\it Bull. Seis. Soc. Am., 70}, 1607-1633, 1980.

\reference Suyehiro, S., Difference between aftershocks and foreshocks in the
	relationship of magnitude frequency of occurrence for the great chilean
	earthquake of 1960,   {\it Bull. Seis. Soc. Am., 56}, 185-200, 1966.

\reference  Tajima, F. and H. Kanamori, Global survey of aftershock area
	expansion patterns, {\it Phys. Earth Planet. Inter., 40}, 77-134, 1985.

\reference Utsu, T., Y. Ogata and S. Matsu'ura, The centenary of the
           Omori Formula for a decay law of aftershock activity,
           {\it J. Phys. Earth}, {\it 43}, 1-33, 1995.

\reference Wu, K.T., Yue, M.S., Wu, H.Y., Chao, S.L., Chen, H.T.,
Huang, W.Q., Tien, K.Y and S.D. Lu,
Certain characteristics of the Haicheng earthquake (M=7.3) sequence,
    {\it Chinese Geophysics, AGU, 1}, 289-308, 1978.

\reference Wyss, M. and W. H. K. Lee, Time variations of the average
	magnitude in central California,
	{\it Proceedings of the conference on Tectonic Problems of
	the San Andreas Fault System}, edited by R. Kocach and A. Nur,
	Stanford University Geol. Sci., 13, 24-42, 1973.

\reference Yamanaka, Y. and K. Shimazaki, Scaling relationship between
             the number of aftershocks and the size of the mainshock,
              {\it J. Phys. Earth, 38}, 305-324, 1990.

\reference Yamashina, K., Some empirical rules on foreshocks and earthquake
	prediction, in {\it Earthquake Prediction}, edited by D. W. Simpson
	and P. G. Richards, AGU Maurice Ewing series: 4, Washington, D.C., pp
	517-526, 1981.

\reference Yamashita, T. and L. Knopoff,
	Models of aftershock occurrence,
	{\it Geophys. J. R. Astron. Soc.}, {\it 91}, 13-26, 1987.

\reference Yamashita, T. and L. Knopoff,
          A model of foreshock occurrence.
        {\it Geophys. Journal}, {\it 96}, 389-399, 1989.

\reference Yamashita, T. and  L. Knopoff,
	Model for intermediate-term precursory clustering of earthquakes,
	{\it J. Geophys. Res.}, {\it 97}, 19873-19879, 1992.

\reference Zhang G., Zhu L., Song, X., Li, Z., Yang, M., SU, N., X. Chen,
	Predictions of the 1997 strong earthquakes in Jiashi, Xinjiang, China,
       {\it Bull. Seis. Soc. Am., 89}, 1171-1183, 1999.

\reference Zoller, G. and S. Hainzl,
	A systematic spatio-temporal test of the critical point hypothesis
	for large earthquakes, {\it Geophys. Res. Lett.}, accepted, 2002.

\end{references}
\end{document}